# Genesis of primitive Hawaiian rejuvenated-stage lavas: Evidence for carbonatite metasomatism and implications for ancient eclogite source

**Anastassia Y. Borisova** [1,2*]

[1] Géosciences Environnement Toulouse, GET - UMR 5563 - OMP - CNRS, 14 Avenue E. Belin, 31400 Toulouse, France

[2] Geological Department, Lomonosov Moscow State University, MGU, Vorobievu Gori, 119991, Moscow, Russia

*Corresponding author: E-mail: anastassia.borisova@get.omp.eu;

*Corresponding adresse: Géosciences Environnement Toulouse UMR 5563, Observatoire Midi Pyrénées, 14 Avenue E. Belin, 31400 Toulouse, France; Tel: +33(0)5 61 54 26 31; Fax: +33(0)5 61 33 25 60

**Abstract** – To constrain a contribution of deep carbonated mantle, to fractionation of Hf relative to rare earth elements (REE) in volcanic series, we examine available high-quality data on major, trace element and Nd-Hf isotope compositions of ~280 primitive lavas and glasses (MgO = 8.5 – 21 wt%, $SiO_2$ = 37 - 50 wt%) erupted during preshield, postshield and mostly rejuvenated stage of the Hawaiian hot spot (Pacific Ocean). Strong variations of Hf/Sm, Zr/Sm, Ti/Eu, K/Th, Nb/Th, La/K and Ba/K in the lavas are not features of the melt equilibration with residual amphibole or phlogopite, and cannot be due to variable degrees of batch or dynamic melting of uncarbonated lherzolite source. Enrichment in REE, Th and Ba relative to K, Hf, Zr, Ti and Nb together and low Si, high Na, K and Ca contents in the Hawaiian lavas are compositional features of carbonated mantle lithospheric to asthenospheric peridotite source affected by carbonatite metasomatism at temperatures higher than 1100°C and pressures higher than 2 GPa. In contrast, major and trace element signatures of most primitive preshield- and postshield-stage magmas require pyroxenite source.

The available data infer that Salt Lake Crater garnet pyroxenite xenoliths hosted by the Koolau volcano lavas on Oahu, Hawaii, were derived from deep (up to 9 GPa) eclogite source likely generating the carbonatite melts within the Hawaiian plume. These carbonatite melts bear the Salt Lake Crater pyroxenite xenolith series to the surface and are responsible for the carbonatite melt-rock reactions recorded in the xenoliths. Highly radiogenic Hf and decoupled Nd-Hf isotope systematics recorded in the Salt Lake Crater mantle xenolith series on Oahu may be explained by strong Hf fractionation relative to REE owing to ≥1 Ga event of carbonatite metasomatism, which is likely related to partial melting of the deeply subducted carbonated eclogite within the Hawaiian plume.

# INTRODUCTION

Carbonatite melts have been widely documented in oceanic hot spots during several decades (e.g., Schmidt & Wiedendorfer, 2018 and references therein). Examples are carbonatites of Cape Verde and Canary Islands (Allègre et al., 1971; Silva et al., 1981; Barrera et al., 1981; Gerlach et al., 1988; Hoernle et al., 2002), the Solomon Islands (Nixon and Boyd, 1979), carbonatite melts in xenoliths of the oceanic upper mantle of Canary, Kerguelen Islands and Fernando de Noronha (e.g., Kogarko et al., 1995; 2001; Schiano et al., 1994). Based on composition of peridotite phases and detailed melt inclusion studies, Hauri et al. (1993), Coltorti et al. (1999), Schiano et al. (1994), Mattielli et al. (1999), Hassler (1999), Grégoire et al. (2000), Kogarko et al. (2001) and Hoernle et al. (2002) inferred that widespread ultramafic xenolith series in oceanic environment are affected by carbonatite metasomatism (the Savai'i, Tubuai, Canary, Fernando de Noronha, Grande Comore and Kerguelen Islands). Such a world-wide occurrence of carbonatite features (e.g., Salters & Shimizu, 1988) implies overall geochemical role of the carbonated mantle as a source of the oceanic hot spot magmas. In fact, existence of kimberlite, carbonatite-rich melts and the carbonatite metasomatism in the sources of the Hawaiian lavas has been suspected and recently discussed (e.g., Salters & Zindler, 1995; Keshav & Sen, 2003; 2004; Writh & Rocholl, 2003; Sen et al., 2005; Frezzotti & Peccerillo, 2007; Dixon et al., 2008; Rocholl et al., 2015; Schmidt & Wiedendorfer, 2018 and references therein; Rocholl et al., 2019). Nevertheless, the mostly accepted model of the rejuvenated magma origin is partial melting of garnet-bearing lherzolite leaving residual apatite, phlogopite and titanite as well as melting of mixed lithosphere-plume source composed by lherzolites and pyroxenite veins (e.g., Clague and Frey, 1982; Lassiter et al., 2000; Frey et al., 2000; Sen et al., 2005; Clague et al., 2006; Garcia et al., 2016). In contrast, recent model of Dixon et al. (2008), where depleted mantle at the margins of the plume at sub-lithospheric depths is metasomatised by low degrees partial melts of the Hawaiian plume including both silicate and carbonatitic melts, is developed based on geochemistry of the primitive Kiekie lavas, Niihau. In our work, we synthetize data on major element and Hf/REE, Zr/REE and related incompatible trace element fractionation in ~280 primitive Hawaiian lavas and glasses and available data on the lava-bearing xenoliths, as well as the Hf-Nd isotope systematics of the Hawaiian lavas and the lava-hosted mantle xenoliths.

Hf/REE and Zr/Hf variations in intraplate magmas probably reflect a compositional heterogeneity of their mantle sources caused by carbonatite metasomatism (e.g., Dupuy et al., 1992). However, Hf/REE and Zr/Hf variations in the oceanic island basalts (OIB) and mid-ocean ridge basalts (MORB) could be also controlled by fractional crystallisation of clinopyroxene (e.g., David et al., 2000). Mantle mineralogical composition is proposed to result in the following order of the bulk distribution coefficients of Hf, Sm and Eu between the mantle source and basaltic melts during generation of OIB and MORB: $D_{Sm}<D_{Hf}<D_{Eu}$ (David et al., 2000). However, what is the geochemical tracer of the carbonated mantle source of alkaline series? To examine principal features of carbonatite metasomatism and an essential role of the carbonated peridotite source in rejuvenated-stage Hawaiian magmas, we used an extended database on major and trace element contents in the mostly rejuvenated-stage lavas, but also some primitive lavas erupted during preshield and postshield stages, trace element partitioning between minerals and silicate and carbonatite melts, and trace element compositions of carbonatite magmas, melts and carbonated peridotite xenoliths. We preferred analytical data with well-analyzed Hf, Zr and other incompatible trace element contents and Nd-Hf isotope compositions of the rock and glass samples available in the literature since 1982. The data imply an existence of recycled carbonated eclogite source for the carbonatite melts in the Hawaiian hot spot and mechanisms of Nd-Hf isotope decoupling in the mantle source of the primitive Hawaiian lava series and the related mantle xenoliths.

## 2. SAMPLE DESCRIPTION

**2.1 Samples and analyses selected for this study**

The major criteria of data selection for the database was a high quality of major and trace element analyses, representativity of the analyses for the Hawaiian series and an absence of secondary alteration features in the samples. We selected available major and trace element compositions of ~280 primitive lavas and glasses ($SiO_2 \leq 50$ wt.% and $MgO \geq 8.5$ wt.%) (**Fig. 1**) from the Hawaiian Islands, Pacific Ocean. We preferred high-quality analyses of trace element concentrations obtained using neutron activation or inductively coupled plasma mass spectrometry for Hf, Ta, Th and REE and X-ray fluorescence for Zr, Nb and Y. We used trace element composition of lavas from the Honolulu, Haleakala, Koloa, submarine Wailau landslide, North Arch volcanic field, west Maui, and Mauna Kea volcanoes from Molokai Niihau, Kauai and other Hawaiian islands, being related to mostly rejuvenated and post-shield stages as well as an example of Loihi Seamount, corresponding to the pre-shield stage of volcanic activity (Clague and Frey, 1982; Frey & Clague, 1983; Chen and Frey, 1985; Clague & Dalrymple, 1988; Maaløe et al., 1992; Chen et al., 1991; Reiners and Nelson, 1998; Frey et al. 2000; Dixon & Clague, 2001; Clague & Moore, 2002; Gaffney et al. 2004; Clague et al., 2006; Dixon et al., 2008; Cousens & Clague, 2015; Phillips et al., 2016) (Table 1). Variable analytical techniques in different laboratories were used in the 'old' studies, therefore the data sets are not identical. Nevertheless, based on the previous suggestions from authors (e.g., Clague & Frey, 1982; Frey & Clague, 1983; Clague & Dalrymple, 1988; Chen et al., 1991), we believe that the 'old data' are accurate in the limit of ±10 %. Later works report analytical precision for major element analyses to be better than 0.5%, whereas trace-element analyses performed by inductively coupled plasma mass spectrometry (ICP-MS) and inductively coupled plasma (ICP) methods, have $2\sigma$ precision on trace-element analyses in the limit of 3 to 5% (Reiners & Nelson, 1998). Major element (X-ray fluorescence; XRF) and trace element (ICP-MS) analyzes completed at the Washington State University GeoAnalytical Laboratory demonstrate the 2σ analytical precision (based upon repeat analyses of the BCR-P standard) better than 3%, whereas trace elements were analyzed with precision better than 6 – 7 % for lithosphile element and better than 15% for Ni, Sc, Ga (Gaffney et al., 2004; Dixon et al., 2008).

The primitive preshield-, postshield- and rejuvenated-stage lavas and glasses include ijolites, melilite nephelinites, nepheline melilites, nepheline basanites, alkaline olivine basalts and some transitional-type basalts (**Fig. 1**; Table 1; Supplementary Dataset 1). The alkaline lavas erupted after an interval of erosion lasting 0.25 – 2.5 millions years (Clague, 1987 and references therein). Petrography and major element composition of the lavas are described elsewhere (e.g., Clague and Frey, 1982; Garcia et al., 1986; Clague and Dalrymple, 1988). The rejuvenated-stage lavas are characterized by more depleted isotope characteristics (e.g., lower $^{87}Sr/^{86}Sr$, higher $^{143}Nd/^{144}Nd$, higher $^{176}Hf/^{177}Hf$) relative to shield stage basalts and those of primitive mantle (Stille et al., 1983; Frey et al., 2000; Frey et al., 2005; Fekiacova et al., 2007; Garcia et al., 2010). For example, available Hf-Nd isotope systematics of the rejuvenated-stage lavas suggest the presence of a depleted component within the Hawaiian plume that is distinct from the mantle source of Pacific MORB (Bizimis et al., 2005; 2013).

**2.2. Compositional features of the selected samples**

The selected compositions of ~280 primitive lavas and glasses are limited by concentrations of $SiO_2$ ≤ 50 wt.% and MgO ≥ 8.5 wt.% (**Fig. 1**) from the Hawaiian Islands, Pacific Ocean. The selected samples are characterized by a wide ranges of trace element abundances: Zr (69 - 260 ppm), Hf (1.7 – 6.9 ppm), La (8.5 - 109 ppm), Sm (3.3 - 17 ppm) and Lu (0.05 - 0.61 ppm). To be able to use the database of Reiners and Nelson (1998), where no Hf concentrations are available, we have used available Zr contents and the constant Zr/Hf = 36 ratio corresponding to that of the primitive mantle after Sun & McDonough (1989); and the Hf contents have been calculated according to the Zr contents for this lava series. **Figure 2** highlights trace element ratios of the Hawaiian series. The investigated Hawaiian series show similar trace element ratios and primitive mantle-normalized patterns. These series exhibit positive correlations between $(Hf/Sm)_n$ from 0.2 to 1.1 (all ratios are normalized to the primitive mantle composition of Sun and McDonough (1989)) and $(Ti/Eu)_n$ varying from 0.3 to 1.2. Zr/Hf ratios vary from 43 to 64 in the Hawaiian series (**Fig. 2**). $(Hf/Sm)_n$ (or $(Zr/Sm)_n$) with $(La/K)_n$ ranging from 1.0 to 10 exhibit a negative correlation, reflecting coherent K, Hf (or Zr) and Ti depletion relative to REE compared to the composition of the primitive mantle. In the Hawaiian rejuvenated-stage series, positive correlations between $(Hf/Sm)_n$ (or $(Zr/Sm)_n$) and $(K/Th)_n$ (0.1-1.4), and between $(Hf/Sm)_n$ (or $(Zr/Sm)_n$) and $(Nb/Th)_n$ (0.6-2.6) reflect that the degrees of the K and Nb depletions relative to Th correlate with intensity of Hf (or Zr) depletion relative to REE. In the lava series, $(Ba/K)_n$ vary widely from 2 to 12, implying an absence of coherent K and Ba behavior. Broad negative correlation of $(Ba/K)_n$ with $(Hf/Sm)_n$ in the lava series suggest that the most Hf depleted lavas and glasses are also characterized by Ba enrichment relative to K (**Fig. 2**).

The Hawaiian rejuvenated-stage series exhibit negative correlations between $(Hf/Sm)_n$, and $(La/Sm)_n$, reflecting coherent light rare earth element (LREE) enrichment in the lavas most depleted in Hf relative to REE. The primitive mantle-normalized trace element patterns of the rejuvenated-stage lavas having low $(Hf/Sm)_n$ (<1) and high $(La/Sm)_n$ show strong negative Rb, K, Zr, Hf and Ti and positive Ba and Th anomalies (**Fig. 3**). **Figure 3** demonstrates that the primitive mantle-normalized patterns of postshield-stage lavas with highest $(Hf/Sm)_n$ (~1) and low $(La/Sm)_n$ exhibit a weak negative anomaly of K and an absence of Zr, Hf and Ti anomalies. The both patterns show different degrees of Rb and Th depletion relative to LREE.

# 3. RESULTS AND DISCUSSION

To constrain geochemical significance of the observed Hf/REE fractionation and related trace element fractionation in the Hawaiian volcanic series, we have considered several possible mechanisms.

## 3.1. Effect of crystallisation on trace element composition

Ni content (200 – 600 ppm) evidences broad positive correlation with MgO contents in most lava series, reflecting mixed signals of olivine fractionation or accumulation, whereas an absence of any correlation of Sc (10 – 30 ppm) with MgO (broadly constant Sc content for Hawaiian series) exclude any significant influence of either clinopyroxene crystallisation or clinopyroxene accumulation (**Figs. A1, A2**).

## 3.2. Amphibole or phlogopite as residual phases

Trace element contents of oceanic island basaltic melts may be controlled by partial melting of hydrated mantle in the presence of residual amphibole (Amph) or phlogopite (Phl) (e.g., Clague and Frey, 1982; Class and Goldstein, 1997; Class et al., 1998; Frey et al., 2000). Both Amph and Phl show similar patterns of partition coefficients between the minerals and basanite melts (Green, 1994; LaTourette et al., 1995). The principal features of Amph and Phl are enrichment in K, Nb, Th, Ba and Ti relative to REE (**Table 2**, LaTourette et al., 1995; Brenan et al., 1995). If Amph and Phl were responsible for K and Ti depletion in the magmas, then magmas would have exhibited a corresponding Ba depletion (e.g., La Tourette et al., 1995). **Figure 2** demonstrates that if Amph were residual mineral phase during partial melting of lherzolite assemblage, Hf/Sm and Ba/K of partial melts would have exhibited positive correlation. Whereas, most of the Hawaiian lava series show negative correlation of $(Hf/Sm)_n$ with $(Ba/K)_n$, expressed in coherent Hf and K depletion relative to REE and Ba that excludes the equilibration of the melts with residual amphibole or phlogopite.

It should be noted that according to the data of Adam et al. (1993), Green (1994) and LaTourette et al. (1995), Hf partition coefficients between Amph and basanite melt and Phl and basanite melt ($K_d^{Hf}$ = 0.008 – 0.33) in both cases overlap those of Sm (0.017 – 0.66) and Zr (0.091 – 0.45) (**Table 2**). Therefore, a decrease in degree of partial melting of mantle source containing Amph and Phl would result in approximately constant Hf/Sm, Zr/Sm and Zr/Hf in the residual melts relative to the mantle source (**Fig. 2**). Partitioning of La for both phases are much lower than that of K (**Table 2**). Therefore, if Amph or Phl were present as mantle phases during partial melting of a homogeneous mantle source, Hf/Sm (or Zr/Sm) and La/K in equilibrated melts would have shown weak positive correlation. In contrast, the Hawaiian lava series demonstrate clear negative correlation of $(Hf/Sm)_n$ (or $(Zr/Sm)_n$) with $(La/K)_n$ (**Fig. 2**), expressed in coherent Hf, Zr and K depletion relative to REE in the primitive mantle-normalized patterns.

High partitioning of K relative to those of Nb and Th (**Table 2**, LaTourette et al., 1995; Green, 1994) suggests that if Amph and Phl were present during melting of a homogeneous mantle source, Hf/Sm (or Zr/Sm) would have exhibited a weak negative correlation with K/Th and Nb/Th (**Figs. 2, 3**). In contrast, most Hawaiian series $(Hf/Sm)_n$ (or $(Zr/Sm)_n$) show positive correlations with $(K/Th)_n$ and $(Nb/Th)_n$. Therefore, a melting process different from partial melting of Amph- or Phl-bearing peridotite source is required to explain the coherent K, Hf, Zr, Ti depletions relative to Th, Ba and REE in most volcanic series.

### 3.3. Batch and dynamic melting of anhydrous lherzolite mantle

High MgO contents 8.5 – 21 wt.% of the Hawaiian lavas series suggest that the compositions are probably not far from the primary melt composition (e.g., melt in equilibrium with mantle source). The effect of fractional crystallization is negligible and cannot account for strong trace element variations. Experimental data on trace element partition coefficients ($K_d$, **Table 2**) between the main mantle mineral phases (olivine, orthopyroxene, clinopyroxene, spinel, garnet) and basaltic/basanite melt show that: 1) $K_d^{Hf}$ and $K_d^{Sm}$ between mantle minerals and basaltic/basanite melt are similar; 2) $K_d^{Zr}$ between clinopyroxene and melt (0.09 - 0.27) are similar to those of $K_d^{Hf}$ (0.1 - 0.55) and $K_d^{Sm}$ (0.09 - 0.67); 3) $K_d$ depend on temperature (T), pressure (P) and melt composition (X), however, the sequence $K_d^{Zr} \leq K_d^{Hf} \approx K_d^{Sm}$ is observed for clinopyroxene, garnet and other silicate minerals in a wide range of these P-T-X parameters (**Table 2**). For example, partitioning between clinopyroxene and basaltic melt of Hart and Dunn (1993) and Blundy et al. (1998) for Zr, Hf and Sm is the following $K_d^{Zr} < K_d^{Hf} \approx K_d^{Sm}$.

To check an influence of partial melting of anhydrous spinel lherzolite and garnet lherzolite on element fractionation in primary melts, estimates of Hf/Sm, Zr/Sm, Zr/Hf, etc. of the melts relative to those of spinel and garnet lherzolite sources using batch melting model are performed (**Fig. 2**). The estimates suggest that Hf/Sm and Zr/Sm of the partial melts increase slightly while La/K, K/Th and Nb/Th slightly decrease relative to those of the lherzolite source with decreasing melting degrees. Batch melting of the spinel and garnet lherzolite results in weak K, Zr and Hf enrichment in the partial melts relative to the residual lherzolite. By contrast, most rejuvenated Hawaiian lava series demonstrate strong fractionation of trace elements reflecting variable K, Hf, Zr, Ti depletions relative to REE and low Hf/Sm, Zr/Sm, Ti/Eu and high La/K relative to the primitive mantle (**Fig. 2**). It is now considered that partial melting of mantle peridotite is a dynamic process, involving differential flow of melt and residual matrix. Channel-segregation melting and percolation melting models were suggested to account for dynamic melting processes (e.g., Eggins, 1992). However, trace elements with similar bulk distribution coefficients (e.g., Dy and Yb) between the main mantle minerals and partial melts generated during dynamic melting demonstrate an absence of fractionation relative to that predicted by the batch melting model (Eggins, 1992). Additionally, comparison of incongruent dynamic melting with congruent dynamic melting shows that the differences are noticeable when the distribution coefficient of the produced mineral is sufficiently different from those of the reaction minerals (Zou & Reid, 2001). Therefore, dynamic melting of anhydrous lherzolite assemblages cannot cause any strong fractionation of trace elements in the generated melts in the case of similarity of the bulk distribution coefficients for the main mantle minerals (e.g., Hf and Sm, Zr and Sm, La and K, K and Ba, Ti and Eu) (**Fig. 2**). The observed strong trace element variations in the alkaline melts relative to the primitive mantle composition could not be due to neither batch melting of homogeneous anhydrous spinel or garnet lherzolite sources nor dynamic melting involving the residual mantle. Since Amph and Phl are unlikely to be residual phases; and the Hf/Sm, Zr/Sm, Ti/Eu, Ba/K, La/K, K/Th and Nb/Th in the primitive melts depend weakly on melting degree and model of anhydrous lherzolite melting, these ratios in most Hawaiian lavas therefore correspond to those of the mantle sources. The Hf/REE and associated trace element fractionation in the melts could be due to heterogeneities of their mantle source or to different mantle sources with variable Hf/Sm, Zr/Sm, Ti/Eu, Ba/K, La/K, K/Th and Nb/Th or both. The variations of the trace element ratios in most volcanic series seem to result from various degrees of REE, Ba and Th enrichment relative to K, Hf, Zr, Ti and Nb in their mantle sources. Therefore, the trace element fractionation in the Hawaiian series demonstrates strong influence of metasomatic processes.

## 3.4. Effect of carbonatite metasomatism on mantle chemistry

The data and correlations represented in our work support conclusion of Dupuy et al. (1992) that the variations of Hf/Sm, Zr/Hf and Zr/Sm in the oceanic island lavas could be due to carbonatite metasomatism of their mantle source. Additionnally, model of Dixon et al. (2008) introduces carbonatite metasomatism in the source of the primitive Kiekie lavas, Niihau, Hawaii. To reconsider this conclusion according to all available database on the primitive lavas of the Hawaiian series, we used three approaches to estimate the geochemical effects of carbonatite metasomatism on the upper mantle: 1) based on composition of oceanic and continental carbonatite melts and magmas; 2) based on experimental data on the elemental partition coefficients between mantle minerals and carbonatite melts and 3) based on composition of mantle peridotites affected by carbonatite metasomatism.

The general geochemical features of differentiated carbonatite melts are: a) high Ba, Th, Sr, LREE contents relatively to Rb, K, Zr, Hf, Ti and HREE abundances (Nelson et al., 1988; Walter et al., 2008). These features may be explained by different extent of trace element complexation with $CO_3^{2-}$; b) variable Sr, Ta, Nb and P contents and significant Zr/Hf fractionation may be explained by fractional crystallization of carbonatite melt (Nelson et al., 1988; Eggler, 1989; Walter et al., 2008). Data of Nelson et al. (1988), Gerlach et al. (1988), Woolley et al. (1991), Beccaluva et al. (1992) and Tayoda et al. (1994) imply that Zr/Hf vary from 5 to 13,000 in continental and oceanic carbonatite series (**Supplemenray Data 2**). In the carbonated peridotites Zr/Hf also varies strongly. For example, Zr/Hf in the Tanzanian mantle xenoliths vary from 20 to 100 (Rudnick et al., 1993). Since the Zr/Hf in the primitive mantle was estimated to be 36 (Jochum et al., 1989), then effect of the carbonatite metasomatism (or melting of metasomatised peridotites) could result in either increase or decrease of Zr/Hf in the mantle lithosphere relative to that of the primitive mantle. Thus, this ratio itself, when different from that of the primitive mantle, is also a characteristic feature of the carbonatite influence on the mantle lithosphere composition. Additionally, the main features of carbonatite melts could be high REE, Th and Ba enrichment relative to K, Zr, Hf and Ti, resulting in low Zr/Sm, Hf/Sm, Ti/Eu, K/Th and high La/K, Ba/K. Indeed, according to the data on carbonatite melts ($CO_2$ = 17 - 45%wt.%; $SiO_2$ < 7wt.%) of Nelson et al. (1988), Gerlach et al. (1988), Woolley et al. (1991), Beccaluva et al. (1992), Tayoda et al. (1994) and Hoernle et al (2002), these ratios are lower than 1 and vary widely: $(Ti/Eu)_n$ = 0.0002 - 0.17, $(Zr/Sm)_n$ = 0.001 - 0.8 and $(Hf/Sm)_n$ = 0.001 - 0.5, $(K/Th)_n$ = 0.001 - 0.1, whereas $(La/K)_n$ = 3 - 29,898, and $(Ba/K)_n$ = 6 - 3500 are commonly high (normalized to primitive mantle composition of Sun and McDonough, 1989) (**Fig. 4**). In high-silica carbonatite melts (with $SiO_2$ = 7 – 22 wt.%) these ratios vary in the following ranges: $(Ti/Eu)_n$ = 0.2 - 0.3, $(Zr/Sm)_n$ = 0.4 - 0.7, $(Hf/Sm)_n$ = 0.6 - 0.9, $(La/K)_n$ = 3 - 38 and $(Ba/K)_n$ = 4 - 34.

Several experimental investigations of partition coefficients between mantle minerals and carbonatite melts have been performed. Green and Wallase (1988) inferred that the reaction of primary carbonatite melt with spinel lherzolite produced an increase in large-ion lithophile elements (LILE) without significant decrease in Mg/(Mg+Fe). Brenan and Watson (1991) investigated partitioning of trace elements between clinopyroxene and olivine and carbonatite melt. They established that carbonatite melt interaction with depleted lherzolite may markedly raise levels of LILE in clinopyroxene. Based on experimentally determined trace element partitioning between clinopyroxene and carbonatite melt, Klemme et al. (1995) established that the most sensitive indicators of carbonatite metasomatism should be low Ti/Eu in the metasomatised peridotite. Sweeney et al. (1995) investigated trace element partitioning between the most common mantle minerals and carbonatite melt. They concluded that effect of carbonatite metasomatism will result in Ti depletion, and increase in LREE relative to heavy rare earth elements (HREE), resulting in increasing LREE/Hf, LREE/Ti in the residual mantle minerals. Therefore, low Hf/REE and Ti/REE of metasomatised peridotites are typical features of the carbonatite metasomatism.

From available data on metasomatised peridotite xenoliths, we prefered high quality ICP-MS data on trace element concentrations including Hf, Zr, REE, Th, Nb, Ta, etc.. Hauri et al. (1993) and Coltorti et al. (1999) demonstrated that peridotite xenoliths from islands of Savai'i, Tubuai and Grande Comore were affected by carbonatite melts (**Supplementary Dataset 2**). The peridotite xenoliths show high REE abundances relative to Zr and Ti. Ionov et al. (1993) performed a detailed investigation of Spitsbergen peridotite xenoliths containing primary carbonate aggregates and quenched dolomite melts. The Spitsbergen metasomatised peridotite xenoliths show marked enrichment in LREE, Sr, Ba and depletion in Zr, Hf, Nb and Ta. In their investigation of the peridotite xenoliths from western Victoria, Yaxley et al. (1991) described apatite-bearing wehrlites and lherzolites and concluded that an interaction between mantle peridotite and carbonatite melt resulted in LILE and CaO enrichment of these peridotites (Yaxley et al., 1991). Similar features have been described by Rudnick et al. (1993) for the Tanzanian peridotites xenoliths. The main geochemical characteristics of such xenoliths are LREE and REE enrichment and strong Ti depletion relative to Eu. **Figure 4** demonstrates trace element signatures of carbonatite melts and peridotite xenoliths affected by carbonatite metasomatism. The peridotite xenolith compositions described by Ionov et al. (1993), Rudnick et al. (1993) and Yaxley et al. (1991) exhibit low $(Ti/Eu)_n$ = 0.02 - 0.4, $(Hf/Sm)_n$ = 0.1 - 0.7, $(Zr/Sm)_n$ = 0.12 - 1.5, $(K/Th)_n$ = 0.04 - 1.3, $(Nb/Th)_n$ = 0.27 - 1.0 and high $(La/K)_n$ = 1.8 - 54 and $(Ba/K)_n$ =1.2 - 15.8 (normalized to primitive mantle composition of Sun and McDonough, 1989). These ratios reflect variable K, Hf, Zr, Ti and Nb depletion relative to adjacent REE, Th and Ba in the mantle xenoliths affected by carbonatite melts. $(Sm/Nd)_n$ and $(Rb/Sr)_n$ versus $(Hf/Sm)_n$ show negative correlations (**Fig. 4**) in the metasomatised peridotite xenoliths of Rudnick et al. (1993), Ionov et al. (1993) and Yaxley et al. (1991). Therefore, during metasomatic process Rb/Sr and Sm/Nd ratios decrease, thus, reducing growth of Sr and Nd isotope ratios in the mantle source, in accordance to conclusion of Meen et al. (1989). In summary, the main geochemical effect of carbonatite metasomatism is enrichment of upper mantle in REE, LREE, Th and Ba relative to K, Ti, Zr, Hf and Nb, resulting in increase of La/K, Ba/K and decrease in Hf/Sm, Zr/Sm, Ti/Eu, K/Th, Nb/Th relative to the primitive mantle. Since high La/K, Ba/K and low Hf/Sm, Zr/Sm, Ti/Eu in the mantle are typical signatures of carbonatite metasomatism and these ratios in partial melts only slightly depend on degrees of melting of anhydrous mantle lherzolite, then one can conclude that the primitive Hawaiian rejuvenated-stage lavas having high La/K, Ba/K and low Hf/Sm, Zr/Sm, and Ti/Eu were in equilibrium or near-equilibrium with a mantle affected by carbonatite metasomatism.

### 3.5. Chromatographic effect of melt percolation in the lithosphere

Petrogenesis of the comagmatic alkaline and transitional melts from the oceanic island series could be explained by chromatographic reactions of the melts with solid mantle during their migration and ascent (e.g., Alibert et al., 1983; Frey and Roden, 1987; Navon and Stolper, 1987; Hauri, 1997). In this case, trace element composition of the melts reflect partial equilibration of the melts with large volumes of mantle outside their source region. According to Navon and Stolper (1987), this effect is controlled by matrix-melt partition coefficients. The first melts emerging from a chromatographic column are in equilibrium with the matrix and the melt trace element concentrations are similar to those produced by partial melting of the column matrix. **Figure 5** demonstrates that the preshield- and postshield-stage lavas with higher $(Hf/Sm)_n$ show lower alkalinity and higher $SiO_2$ contents, whereas, the rejuvenated-stage lavas with lower $(Hf/Sm)_n$ display higher alkalinity and lower $SiO_2$ contents. Since Hawaiian lavas with low $(Hf/Sm)_n$ are suggested to be in equilibrium with mantle peridotite affected by carbonatite melts, the low $SiO_2$ lavas should be the first strongly modified melts emerging from the carbonated mantle column. The high $SiO_2$ and high $(Hf/Sm)_n$ lavas must be, in the context of this scenario, the late unmodified plume melt. In contrast, high $SiO_2$ lavas which have been described among the postshield-stage lavas (which preceds the rejuvenated-stage lava), having high $(Hf/Sm)_n$ (1.00 - 1.03) show trace element composition and isotopic ratios of $^{87}Sr/^{86}Sr$ (0.7036 - 0.7037) and $^{143}Nd/^{144}Nd$ (0.51295 - 0.51297) (Chen et al., 1991; Hofmann and Jochum, 1996) similar to those of the Hawaiian tholeiites (Figs. 2, 5) considered as the Hawaiian plume-derived magmas (Chen et al., 1991) and not related to the rejuvenated-stage lavas. The most important features of the chromatographic effet of melt percolation is an abrupt change in trace element patterns each time the concentration front reaches the top of the column (e.g., Navon and Stolper, 1987). Correlations of Hf/Sm with $SiO_2$ and $(La/Sm)_n$ imply possible effect of the chromatographic effet of the alkaline melts percolated in the carbonated mantle. Nevertheless, Reiners and Nelson (1998) demonstrated that the isotope-trace element compositions of the rejuvenated-stage lava series are well correlated, whereas melt-mantle interaction model of Hauri (1997) predicts strong decoupling of isotope-trace element composition if the reactive mantle would have different isotopic composition than the melt. Therefore, trace and major element fractionation in the Hawaiian lavas does not confirm that the primitive Hawaiian alkaline lava series might be affected by the chromatographic effect of the melt percolation or melt-rock reaction in the carbonated mantle.

**3.6. Carbonatite metasomatism influence on Hawaiian mantle**

Experimental studies infer that carbonatite melts occur within the mantle lithosphere but may be also stable in the hotter asthenospheric mantle. The experimental work of Green and Wallace (1988) defines distinctive window in oceanic or young continental lithosphere geotherm, in which a sodic, dolomitic carbonatite melts may occur and may cause mantle metasomatism at pressures lower than 3.1 GPa, 930 - 1080°C for fertile lherzolite. Falloon & Green (1989) and Baker and Wyllie (1992) concluded that at pressures higher than 2.1 - 2.2 GPa below 1050°C, near-solidus melts from peridotite + $CO_2$ ± $H_2O$ are dolomitic (Mg-rich carbonatite) melts. Dalton and Wood (1993) experimentally determined that near-solidus melts from depleted natural lherzolite at pressures greater than 2.5 GPa are carbonatite with low alkali contents similar to natural magnesio-carbonatites. Sweeney (1994) demonstrated that dolomitic melts can be in equilibrium with a peridotitic mantle in the following conditions of the upper mantle: 2.0 - 5.0 GPa and 950 - 1200°C. If melt segregation and aggregation occur within the oceanic lithosphere in the range of pressures 2.0 - 5.0 GPa and temperatures 930 - 1200°C, then this melt may move rapidly to the surface within narrow dykes and veins, yielding primary sodic, dolomitic carbonatite magmas or fractionated natro-carbonatite derivatives (Green and Wallace, 1988; Sweeney, 1994). However, if the carbonatite melts infiltrate lherzolite at pressures lower than 2.0 GPa at temperatures from 950 to 1050°C, decarbonation reactions occur and release $CO_2$ (Green and Wallace, 1988; Harmer and Gittins, 1998). Moreover, recent compilation of experimental data infer that the carbonatite melts may be stable at pressure-temperature conditions higher than 2 GPa pressure and 1100°C (Hammouda, 2003) (**Fig. 6**).

Hawaiian mantle xenoliths demonstrate evidence for carbonatite metasomatism influence on the oceanic mantle. Mantle minerals from anhydrous spinel lherzolites of Salt Lake Crater, Honolulu, Hawaii include a dense supercritical $CO_2$ fluid phase (density 1.16 g/cm$^3$, De Vivo et al., 1988) and LREE-enriched clinopyroxene (Salters and Zindler, 1995). Pyroxenite xenoliths from the Salt Lake Crater also consist of newly-formed LREE enriched clinopyroxene (spinel + olivine ± garnet ± orthopyroxene ± amphibole ± phlogopite) containing superdense $CO_2$ (1.21 g/cm$^3$) with apatite (Frezzotti et al., 1992; Sen et al., 1993). The estimated pressures of the fluid inclusion entrapment in the Salt Lake Crater peridotite minerals are lower than 2.0 GPa (Frezzotti et al., 1992). At these pressures (≤ 2.0 GPa) carbonatite metasomatism results in LILE and phosphorus enrichment and conversion of spinel lherzolite to apatite-bearing wehrlite containing olivine and diopside (Green and Wallace, 1988), therefore, decarbonation reaction features are evident in the mantle xenoliths from the Salt Lake Crater. Low Hf/Sm and Ti/Eu in clinopyroxenes of the anhydrous spinel lherzolites (Salters and Zindler, 1995) and in the Samoa harzburgites (Hauri, 1997), and low $(Hf/Sm)_n$, $(Zr/Sm)_n$ and $(Ti/Eu)_n$ (down to 0.71, 0.43 and 0.7, respectively) in the type I and II Salt Lake Crater pyroxenites (Frey, 1980) result from a reaction with carbonatite melts. Crystallization of the plume-derived primitive (Mg-rich) basaltic melts produces pyroxenite veins at pressures of 1.0 – 2.0 GPa (e.g., Borisova et al., 2017). However, the carbonatite melts with a $CO_2 >$ 40 wt.% and very low $H_2O$ content (1 – 2 wt.%) are far more effective agents for metasomatism (Green and Wallace, 1988). Salters and Zindler (1995) pointed out that an estimated trace element characteristics of metasomatic agent for the Salt Lake Crater lherzolites are similar to those of carbonatite melt. Therefore, the Hawaiian lithosphere to asthenosphere mantle is likely affected by carbonatite metasomatism.

Carbonatite metasomatism was also inferred for the mantle lithosphere from Canary Islands (Kogarko et al., 1995; Hoernle et al., 2002), Savai'i, Tubuai (Polynesia Islands) and Grande Comore islands (Hauri et al., 1993; Coltorti et al., 1999). Green and Wallace (1988) define that carbonatite melts can cause carbonatite metasomatism in equilibrium with amphibole-bearing lherzolites at pressures lower than 2.0 GPa. Schiano et al. (1994), Mattielli et al. (1999) and Grégoire et al. (2000) inferred an important role of carbonatite and $CO_2$-rich silicate melts as a metasomatic agents in the Kerguelen Archipelago mantle lithosphere, that result in appearance of new-formed clinopyroxene (Mattielli et al. 1999; Grégoire et al., 2000), apatite (Hassler, 1999), phlogopite and amphibole (Grégoire et al., 2000) in the metasomatised peridotites. Despite some evidences for the pargasite megacryst derivation from mantle in equilibrium with carbonatite melt (Harmer & Gittins, 1998), high $CO_2$ and low $H_2O$ activity during the carbonatite metasomatism would result in partial dehydration reaction. This allows suggesting multiple stages of silicate melt metasomatism forming the amphibole-bearing or phlogopite-bearing mantle. This conclusion is in accordance with conclusions of Keshav et al. (2007) that phlogopites from the Salt Lake Crater xenoliths are not related to the formation of the host xenoliths.

## 3.7. Compositional variations in the primitive Hawaiian lavas: plume-lithosphere interaction

Isotopic investigations of Hawaiian primitive alkaline lavas infer generation of the Hawaiian lavas by melting of a mixed lherzolite/pyroxenite sources (e.g., Lassiter et al., 2000). Figures 2 and 5 demonstrate positive correlations of $(Hf/Sm)_n$ with $SiO_2$ and negative correlations with alkalinity $(K_2O+Na_2O)$, CaO and $(La/Sm)_n$ in the Hawaiian Volcanics. The Salt Lake Crater pyroxenites have elevated $(Hf/Sm)_n$ (0.71 - 1.46) which are characteristic of high $SiO_2$ lavas. Therefore, the high-$SiO_2$ melts with high $(Hf/Sm)_n$ approaching 1 are likely to be pyroxenite-derived. Indeed, Lassiter et al. (2000) and Garcia and Presti (1987) inferred that pyroxenite veins give melts with high $SiO_2$ (40 – 47 wt.%), $Al_2O_3$ (10 – 14 wt.%) and low CaO (10 – 12 wt.%), which correspond to the high $(Hf/Sm)_n$ lavas (**Fig. 5**). The most preshield and postshield high-$SiO_2$ melts with low alkalinity, $(La/Sm)_n$ and CaO reflect a higher degree of melting of pyroxenite veins of mafic composition compared to those of the coexisting lherzolite (Reiners and Nelson, 1998; Lassiter et al., 2000). Indeed, lower solidi for the pyroxenites results in a greater melt productivity relative to most peridotites (e.g., Hirschmann & Stolper, 1996; Yang et al., 1998), resulting in significant magma productivity during these stages (Clague, 1987).

In contrast, an addition of carbonate to peridotite results in generation of melts poor in $SiO_2$ (Hirose, 1997), suggesting the most rejuvenated-stage $SiO_2$-poor melts to be derived due to partial melting of carbonated lherzolites. Low $(Hf/Sm)_n$ (0.26 - 0.71) and $SiO_2$ (< 41 wt.%) contents in the most rejuvenated alkaline lavas of Honolulu, Koloa and other volcanoes of Hawaiian Islands suggest that these melts were derived from carbonated lherzolites (**Fig. 5**). Reaction of carbonatite melts with spinel lherzolite produces an increase in Ca/Al and Na/Ca, an increase in LILE elements (Green and Wallace, 1988), in particular REE in the carbonated lherzolite. Decreasing melting degrees of deeper mantle source also result in higher alkalinity and LREE enrichment of partial melts (**Fig. 2**). Therefore, higher CaO, (La/Sm)$_n$ and alkalinity in the low SiO$_2$ melts characteristic for the rejuvenated-stage series may be explained by both the increasing degree of the carbonated metasomatism of the lherzolite source and the decreasing degrees of partial melting of the carbonated lherzolite. Thus, LREE enrichment in the rejuvenated-stage Hawaiian series with coherent light rare earth element (LREE) enrichment in the most depleted in Hf relative to REE are due to combined effect of the low partial melting, high degree of the metasomatized mantle and possible residual garnet in their lherzolite source.

Thus, carbonated lherzolites are a source for rejuvenated-stage SiO$_2$-poor (CaO-rich, (La/Sm)$_n$-rich) melts, whereas the pyroxenites seem to be a source of pre-shield and post-shield-stage SiO$_2$-richer (CaO-poor, (La/Sm)$_n$-poor) basaltic melts. Partial melting of anhydrous lherzolites fractionates slightly Hf/Sm, Zr/Sm, Zr/Hf, Ti/Eu, K/Th, Nb/Th, Ba/K and La/K relative to the residual assemblage. Therefore, low Hf/Sm and corresponding low Zr/Sm, Ti/Eu, K/Th, Nb/Th and high La/K, Ba/K, Zr/Hf in the rejuvenated-stage Hawaiian lavas (**Fig. 2**) reflect REE, Th and Ba enrichment relative to K, Hf, Zr, Nb and Ti and Zr enrichment relative to Hf in their carbonated lherzolite source.

### 3.8. Isotopic effects of carbonatite metasomatism on lithospheric to asthenospheric mantle

We believe that Hawaiian rejuvenated-stage lava series are derived from the lithospheric to asthenospheric mantle, where lithosphere has maximal thickness of 75 – 90 km (e.g., Gurriet, 1987; Liu & Chase 1991; Lassiter et al., 2000; Schmidt & Wiedendorfer, 2018). Figure 7 demonstrates strongest fractionation of Hf/REE and Lu/Hf in the carbonated mantle and carbonatite magmas. Provided that the carbonatite melts are formed in the asthenospheric mantle, are highly mobile and may be percolated upward to the basis of the lithospheric mantle at about of 2 GPa pressure, the effect of the carbonatite melts will be a drastic fractionation of Lu/Hf from ~0.1 to ~10 in the carbonated mantle. Because carbonatite melts are stable in mantle lithosphere at ~90 km, but may be also stable in the hotter mantle, the first predictable effect of the carbonatite metasomatism is the visible decoupling of Nd-Hf isotope systematics in the carbonated mantle. Strongest Lu/Hf fractionation due to mostly REE enrichment is observed only for a limited Sm/Nd fractionation in the carbonated mantle and carbonatite magmas (**Fig. 7**). This should result in production of mantle lithosphere characterized by strongly decoupled Nd-Hf isotope signatures. Nevertheless, because Hf is strongly incompatible in the carbonatite metasomatic melts, carbonatite metasomatism should not change significantly Hf isotope composition of the recently metasomatized mantle. Thus, Hf isotope composition of recently metasomatized mantle source should reflect the initial Hf isotope signatures of the mantle. In contrast, the extremal radiogenic Hf isotope ratios of the Salt Lake Crater xenolihs and decoupled Nd-Hf isotope systematics of these Hawaiian mantle xenolith series may be explained by Hf fractionation relative to REE owing to $\geq 1$ Ga events of carbonatite metasomatism affecting this subducted mantle component (**Fig. 8**). This conclusion is not in contradiction to recent study of Bizimis et al. (2005) that the depleted component with highly radiogenic Hf represents a recycled oceanic lithosphere.

**3.9. Carbonated eclogite source of the carbonatite melts in the Hawaiian plume**

Most works on the Salt Lake Crater xenoliths from Koolau volcano on Oahu, Hawaii demonstrate several important features, suggesting an existence of deep carbonated eclogite source of the Hawaiian carbonatite melts (Sen et al., 1993; Keshav & Sen 2001; 2003; Wirth & Rocholl, 2003; Keshav & Sen, 2004; Keshav et al., 2005; Bizimis et al., 2005; Frezzotti & Peccerillo, 2007; Bizimis & Peslier, 2015). 1) The xenoliths are high-pressure cumulates related to polybaric magma fractionation within the asthenosphere at depths > 100 km (Keshav & Sen, 2003; 2004; Keshav et al., 2007). Additionally, they provide the first evidence of the deepest majorite garnets never observed among the oceanic pyroxenite xenoliths (e.g., Keshav & Sen, 2001) corresponding to 6 – 9 GPa, i.e., down to ~300 km depth. 2) Moreover, an occurrence of nano-diamonds in pyroxene mineral-hosted inclusions of the garnet pyroxenite xenoliths (Frezzotti & Peccerillo, 2007; Writh and Rocholl, 2003) strongly suggests deep eclogite source for the primary carbonatite melts in the Hawaiian hot spot. The fact that some part of the diamonds have been found in the glass of basaltic composition (Writh and Rocholl, 2003) may imply a diamond genesis at 6 GPa pressure and 1300°C, i.e., conditions of diamond stability as well as liquid immiscibility between carbonatite and basaltic melts (Hammouda, 2003). The majorite garnets discovered in the Salt Lake Crater xenoliths are likely derived from eclogite material subducted to the mantle. 3) Radiogenic Hf isotope ratios and decoupled Nd-Hf isotopic systematics of the Salt Lake Crater xenoliths infer that the carbonatite metasomatism affected the lithosphere material stored for more than 1 Ga within the mantle (**Fig. 8**). 4) The "majorite"- derived melts which are fractionated carbonatite melts are enriched in LREE and depleted in high field strength elements (HFSE) (e.g., Walter et al., 2008). 5) The mineralogy and trace element features of the Salt Lake Crater pyroxenite xenolith series also demonstrate evidence for kimberlite veining (Keshav and Sen, 2003) and/or the carbonatite metasomatism (see part 3.6.). 6) Carbonatite-type melts are documented in the Salt Lake Crater lavas (Rocholl et al. 2015; 2019). The conclusion about carbonatite nature of the melts involved in the Hawaiian hot spot is in accordance with evidences for the immisible "kimberlite melts" or carbonate-rich fluids/melts found by Keshav and Sen (2003), Writh and Rocholl (2003), Sen et al. (2005) and Frezzotti & Peccerillo (2007) and for the xenoliths of the Salt Lake Crater series. 7) Finally, model of Dixon et al. (2008) suggests that depleted mantle at the margins of the plume at sub-lithoospheric depths is metasomatised by low degrees partial melts of the Hawaiian plume including both silicate and carbonatitic melts, the conclusion is based on geochemistry of the primitive Kiekie lavas, Niihau. The source of $H_2O$ and $CO_2$ in the Hawaiian glasses and lavas and Salt Lake Crater pyroxenes (Dixon et al., 1997; Dixon & Clague, 2001; Bizimis & Peslier, 2015) may be also due to $H_2O$-bearing carbonatite melts. According to Walter et al. (2008) and experimental model of Hammouda (2003), primary carbonatite melts form due to partial melting of subducted eclogite material within the mantle transition zone and they may crystallize such liquidus minerals as diamonds.

Our model summarized on **Figure 6** demonstrates that recycled carbonated eclogite-derived carbonatite melts are involved in the magmatism of the Hawaiian hot spot. Our model is not in contradiction with the noble gas systematics of the Hawaiian hot spot (Hanyu et al., 2005) and model of carbonated eclogite propagation based on light stable isotope compositions of oceanic basalts (Dixon et al., 2017), although the stable isotope systematics and noble gas geochemistry is beyond the subject of this investigation.

## 4. CONCLUSIONS

1) Primitive alkaline lavas series from the Hawaiian Islands ($SiO_2 \leq 50$ wt.% and $MgO \geq 8.5$ wt.%), erupted during recent rejuvenated stages of the Hawaiian island activity exhibit similar compositional trace element trends. These lavas show REE, Th and Ba enrichments relative to K, Zr, Hf, and Ti, which result in high La/K, Ba/K, Zr/Hf and low Hf/Sm, Zr/Sm, Ti/Eu, and K/Th relative to the primitive mantle composition. Analyses of the available partitioning between amphibole and phlogopite and basanite melts suggest that the trace element composition of the series are not controlled by residual amphibole or phlogopite. Batch or dynamic melting of homogeneous uncarbonated lherzolite mantle cannot explain the coherent K, Hf, Zr and Ti depletion relative to REE in the rejuvenated-stage Hawaiian lavas.

2) The main geochemical effect of carbonatite metasomatism is a strong enrichment of the Hawaiian mantle lithosphere to asthenosphere in rare earth elements, Th and Ba relative to K, Zr,

Hf, Ti and Nb which leads to an increase in La/K, Ba/K and decrease in Ti/Eu, Zr/Sm, Hf/Sm, K/Th and Nb/Th in the carbonated mantle. The primitive Hawaiian lavas with low Hf/Sm, Zr/Sm, Ti/Eu, K/Th and Nb/Th were derived from the lithospheric to asthenospheric peridotite affected by carbonatite metasomatism at temperatures higher than 1100°C and pressures $\geq 2.0$ GPa.

3) The major and trace element composition of the most primitive Hawaiian preshield and postshield-stage lavas demonstrates that high $SiO_2$ (>42 wt.%) contents, $(Hf/Sm)_n$ (>0.71) (and CaO-poor and La/Sm-poor) lavas are high-degree melts derived from pyroxenite veins. In contrast, most rejuvenated-stage lavas with lower $SiO_2$ (<42 wt.%) contents, $(Hf/Sm)_n$ (<0.71) (and CaO-rich and La/Sm-rich) are low-degree melts derived from the deep carbonated lherzolite source. The variations of major and trace element compositions of the primitive lava series of the rejuvenated stage of Hawaiian Islands are well explained by different degree of partial melting and different degree of the carbonatitic metasomatism reaction. This conclusion implies an important role of thermal carbonated lithospheric to asthenospheric mantle affected by carbonatite metasomatism as mantle source of the Hawaiian hot spot.

4) The isotopic, chemical and mineralogic features of the Salt Lake Crater pyroxenite xenoliths from Koolau volcano on Oahu, Hawaii, demonstrate evidences for their derivation from deep (up to 9 GPa) carbonated eclogite source of the carbonatite melts likely involved into the rejuvenated-stage magma origin. Nd-Hf isotope systematics of the carbonated mantle xenolith series may be explained by Hf fractionation relative to REE owing to $\geq 1$ Ga event(s) of carbonatite metasomatism affecting the Hawaiian eclogite source.

*Acknowledgments* - L.V. Dmitriev, F.A. Frey, N. Mattielli, D. Weis, J. Scoates and V. Kamenetsky and M. Grégoire are thanked for very important comments and suggestions, which significantly improved this manuscript.

Class C., Goldstein S.L., Altherr R. and Bachelery P. (1998) The Process of plume-lithosphere interactions in the ocean basin - the case of Grand Comore. *J. Petrol.* **39**, 881-903.

Coltorti M., Bonadiman C., Hinton R.W., Siena F. and Upton B.G.J. (1999) Carbonatite metasomatism of the oceanic upper mantle: evidence from clinopyroxenes and glasses in ultramafic xenoliths of Grande Comore, Indian Ocean. *J. Petrol.* **40**, 133-165.

Cousens B.L., Clague D.A., (2015). Shield to Rejuvenated Stage Volcanism on Kauai and Niihau, Hawaiian Islands. Journal of Petrology, 56, 1547-1584.

Cox K.G., Bell J.D., Pankhurst R.J. (1979). The interpretation of igneous rocks. Allen & Unwin, London, pp. 450.

Dalton J.A. and Wood B.J. (1993) The composition of primary carbonatite melts and their evolution through wallrock reaction in the mantle. *Earth Planet. Sci. Lett.* **119**, 511-525.

David K., Schiano P. and Allègre C.J. (2000) Assessment of Zr/Hf fractionation in oceanic basalts and continental materials during petrogenetic processes. *Earth Planet. Sci. Lett.* **178**, 285-301.

De Vivo B., Frezzotti M. L., Lima A. and Trigila R. (1988) Spinel lherzolite nodules from Oahu island (Hawaii): A fluid inclusion study. *Bull. Mineral.* **111**, 307-319.

Dixon J.E., Clague D.A., Wallace P., Poreda R. (1997). Volatiles in alkalic basalts from the North Arch volcanic field, Hawaii: Extensive degassing of deep submarine-erupted alkalic series lavas. *J. Petrol*. **38**, 911- 939.

Dixon, J. E., and D. A. Clague, (2001). Volatiles in basaltic glasses from Loihi Seamount, Hawaii: Evidence for a relatively dry plume component, *J. Petrol.,* **42**, 627–654.

Dixon J., Clague D.A., Cousens B., Monsalve M.L., Uhl J. (2008). Carbonatite and silicate melt metasomatism of the mantle surrounding the Hawaiian plume: Evidence from volatiles, trace elements, and radiogenic isotopes in rejuvenated-stage lavas from Niihau, Hawaii. *Geochemistry Geophysics Geosystems*, **9**, Q09005, doi:10.1029/2008GC002076.
29

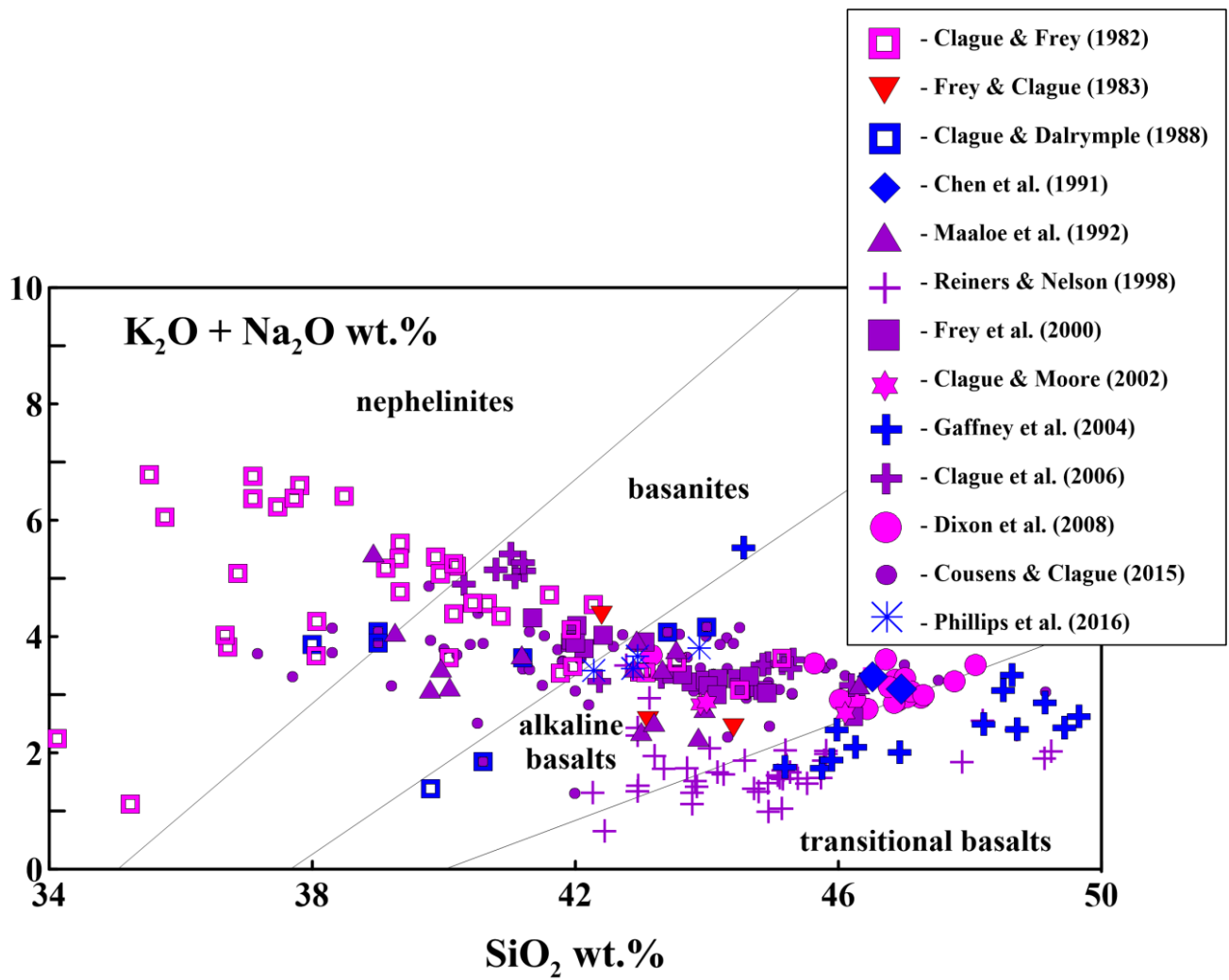

**Figure 1.** K$_2$O+Na$_2$O versus SiO$_2$ contents classification diagram for Hawaiian island lava series (nephelinites, basanites, alkaline olivine basalts, and transitional basalts). The classification is after Cox et al. (1979). The compositional database of the Honolulu, Haleakala, Koloa, submarine Wailau landslide, North Arch volcanic field, west Maui, and Mauna Kea volcanoes from Molokai Niihau, Kauai and other Hawaiian islands, being related to rejuvenated, postshield stage as well as an exemple of Loihi Seamount, corresponding to the preshield stage of volcanic activity (Clague and Frey, 1982; Frey & Clague, 1983; Clague & Dalrymple, 1988; Maaløe et al., 1992; Chen et al., 1991; Reiners and Nelson, 1998; Frey et al. 2000; Clague & Moore, 2002; Gaffney et al. 2004; Clague et al., 2006; Dixon et al., 2008; Phillips et al., 2016) are used on the plot. Red color indicates the preshield stage lavas, blue color corresponds to the mostly postshield stage samples and the purple and magenta colors show the rejuvenated stage lavas and glasses. The data source is given in Table 1 and is summarized in Supplementary Dataset 1.

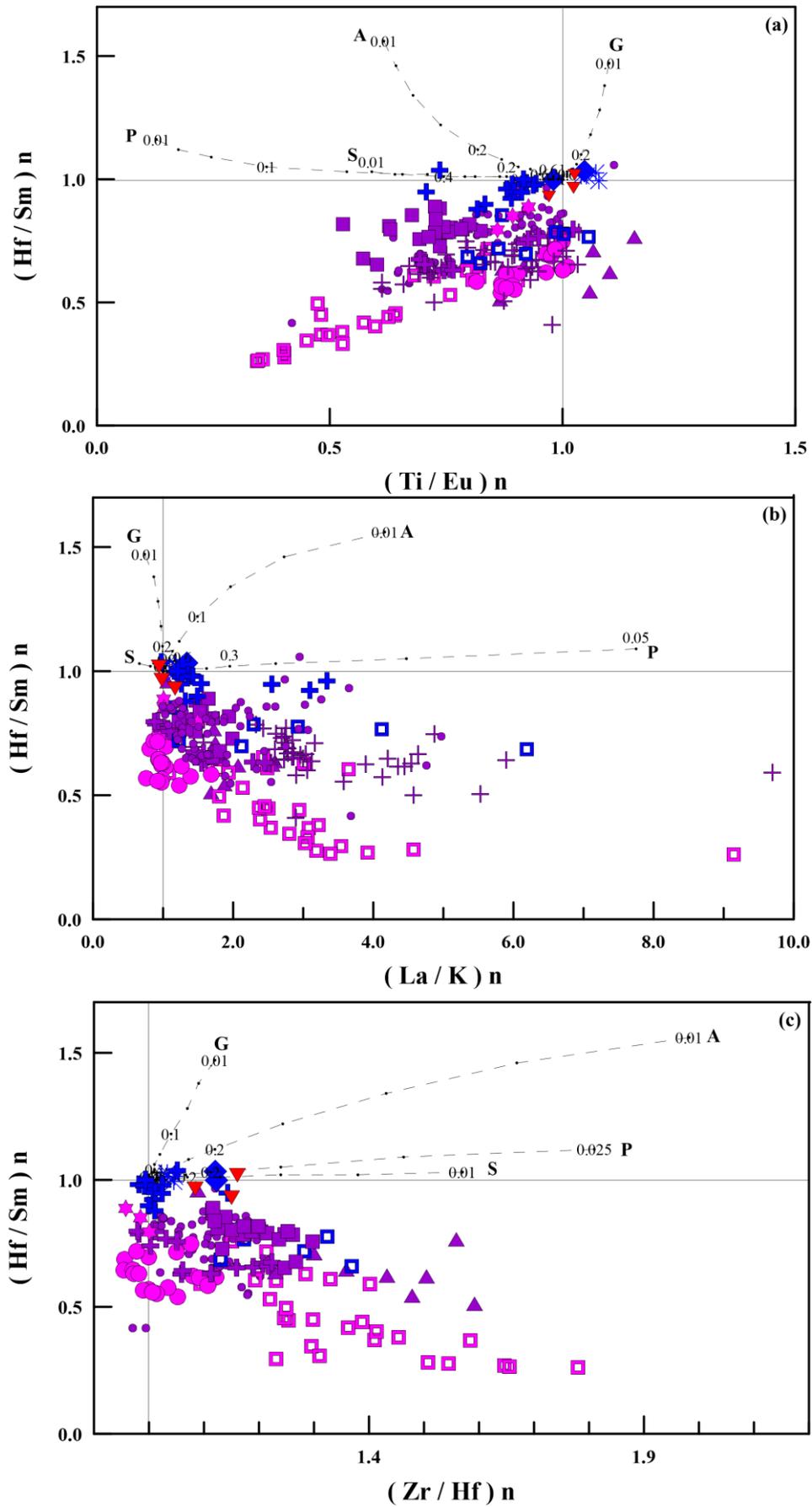

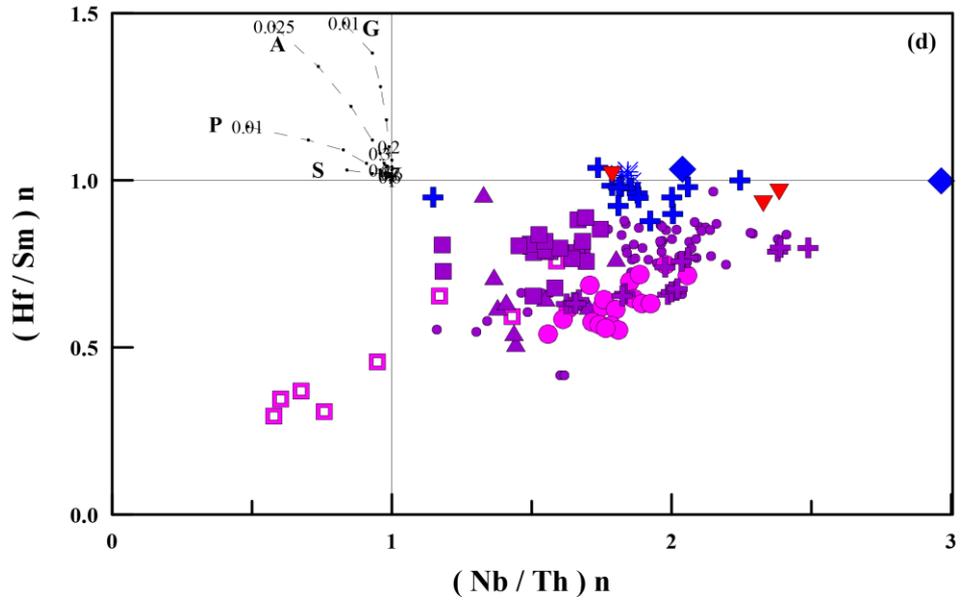

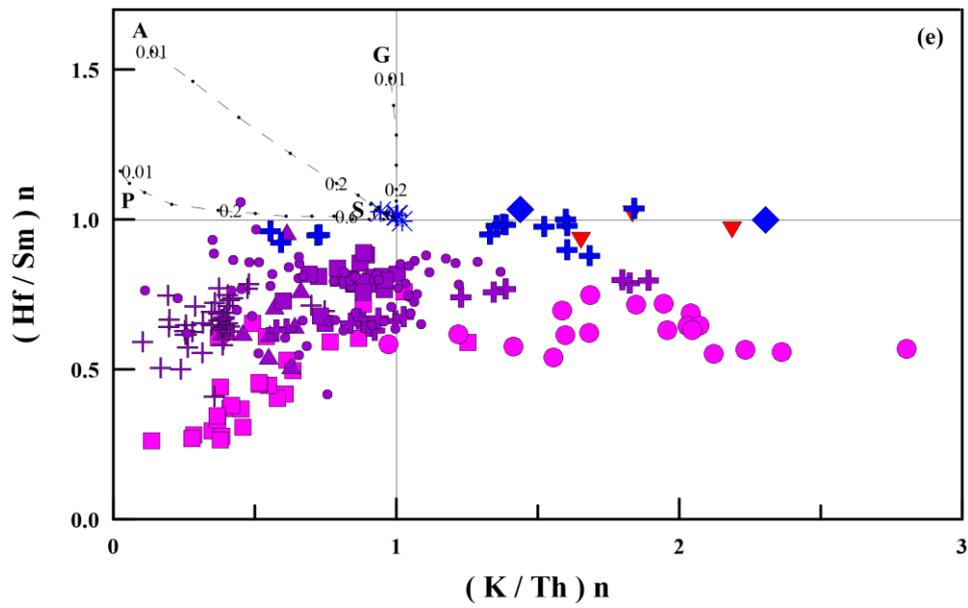

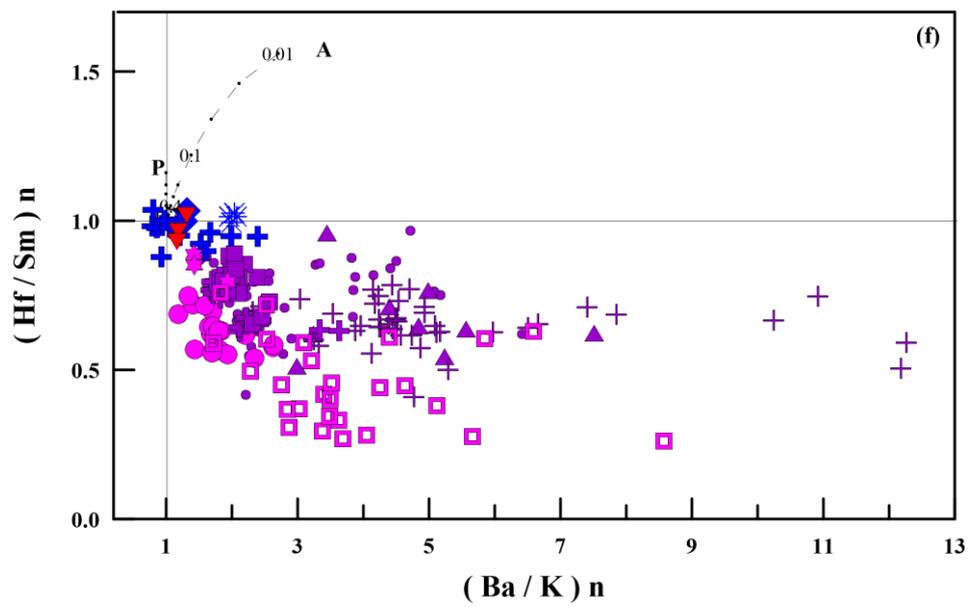

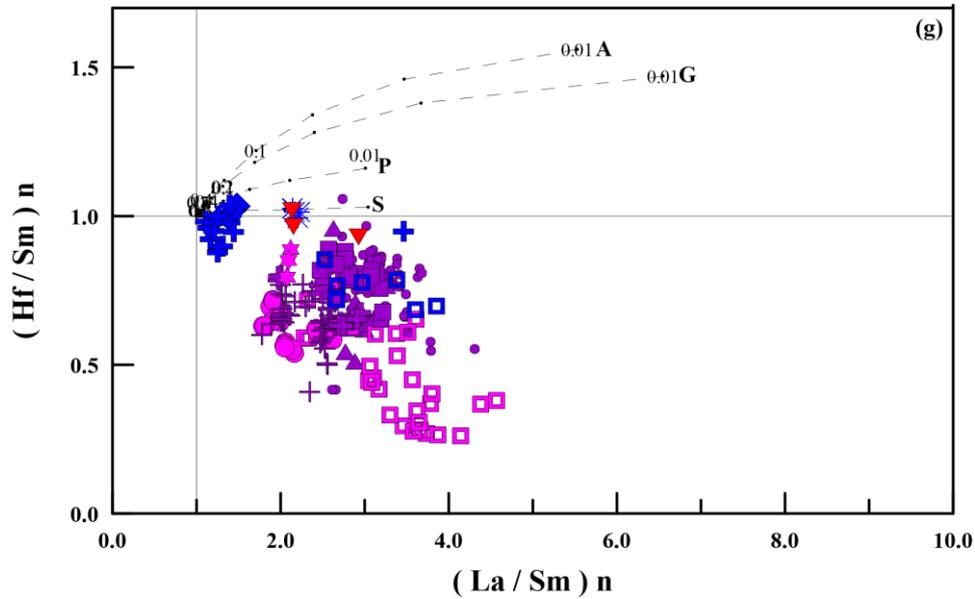

**Figure 2.** Trace element compositions of alkaline Hawaiian island magmas. (a) $(Hf/Sm)_n$ versus $(Ti/Eu)_n$; (b) $(Hf/Sm)_n$ versus $(La/K)_n$ (c) $(Hf/Sm)_n$ versus $(Zr/Hf)_n$ (d) $(Hf/Sm)_n$ versus $(Nb/Th)_n$ (e) $(Hf/Sm)_n$ versus $(K/Th)_n$; (f) $(Hf/Sm)_n$ versus $(Ba/K)_n$; (g) $(Hf/Sm)_n$ versus $(La/Sm)_n$. The used database of the Honolulu, Haleakala, Koloa, submarine Wailau landslide, North Arch volcanic field, west Maui, and Mauna Kea volcanoes from Molokai Niihau, Kauai and other Hawaiian islands, being related to rejuvenated and postshield stages as well as an exemple of Loihi Seamount, corresponding to the preshield stage of volcanic activity. The used data sets are given in Supplementary Dataset 1 and Table 1. Red color indicates the preshield stage lavas, blue color corresponds to the mostly postshield stage samples and the purple and magenta colors show the rejuvenated stage lavas and glasses. The calculated curves demonstrate trace element ratio fractionation in partial silicate melts relative to the mantle sources represented by spinel lherzolite (S), garnet lherzolite (G), amphibole-bearing (A) and phlogopite-bearing (P) lherzolites. Numbers along the curves are corresponding to the fractions of partial melting ($0 < f < 1$).

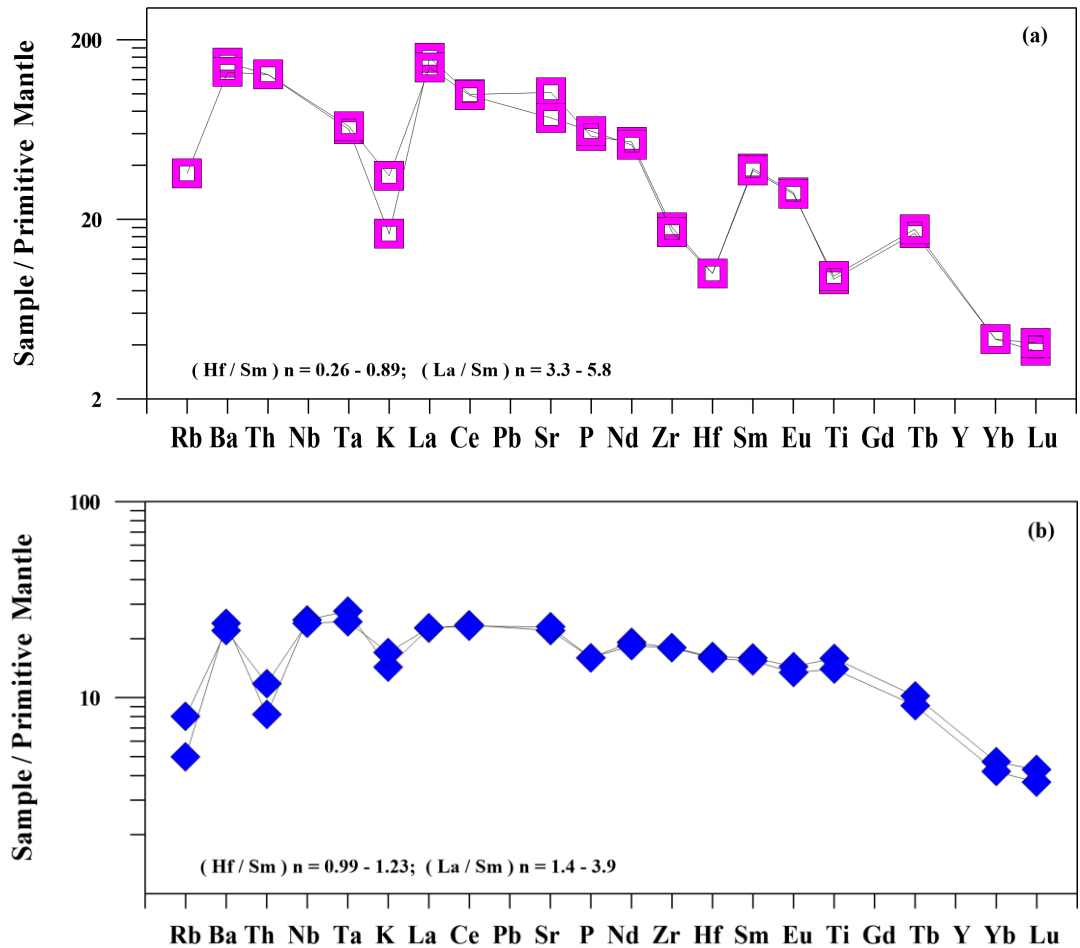

**Figure 3.** Mantle-normalized trace element patterns of Hawaiian lavas with (a) the rejuvenated-stage lava samples (69KAL1 and 69KAL2) with low $(Hf/Sm)_n=0.26-0.89$ and high $(La/Sm)_n=3.3-5.8$ and (b) postshield-stage lava samples (Ho-12 and Ho-14) with high $(Hf/Sm)_n=0.93-1.23$ and low $(La/Sm)_n=1.4-3.9$. Data are from Clague & Frey (1982) and Chen et al. (1991). The patterns are normalized to composition of the primitive mantle according to Sun and McDonough (1989). The data sources are the same as in Figure 1. The data sources are given in Table 1.

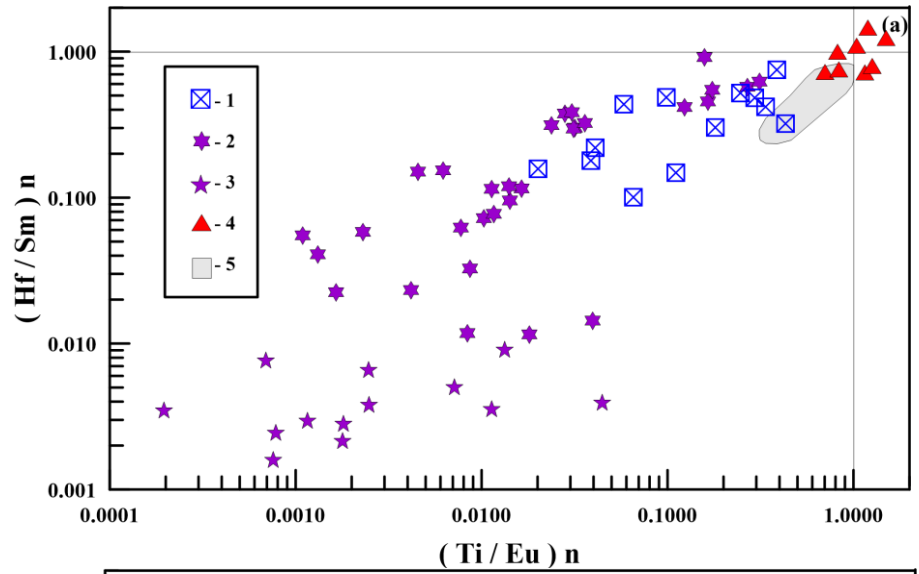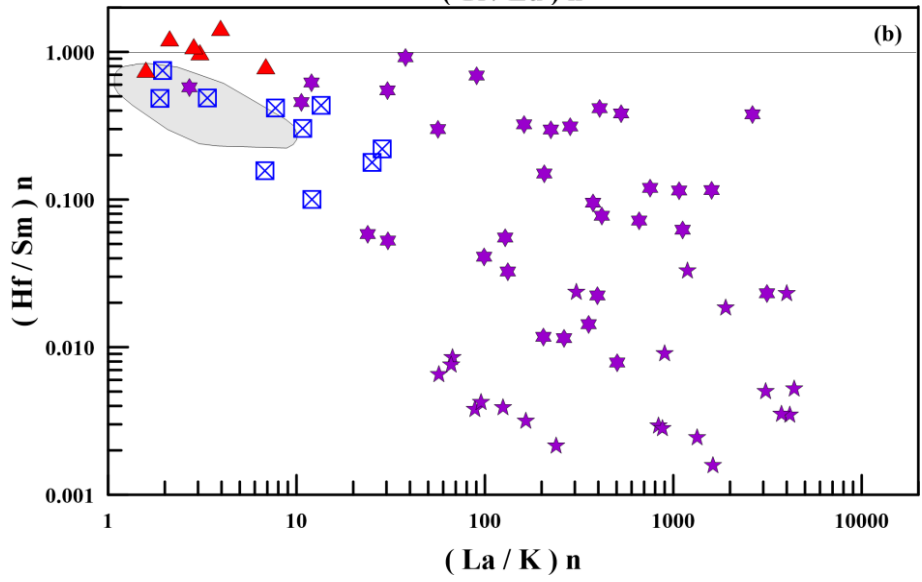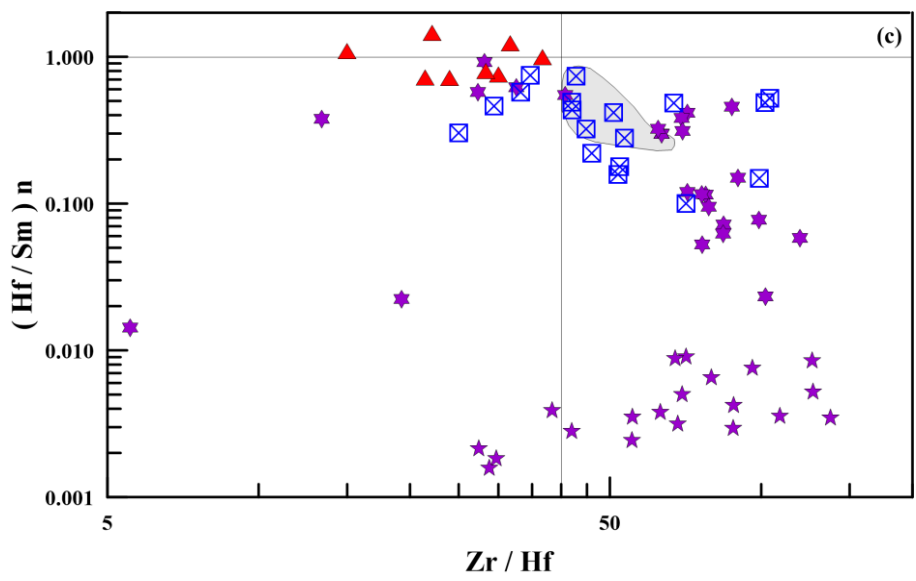

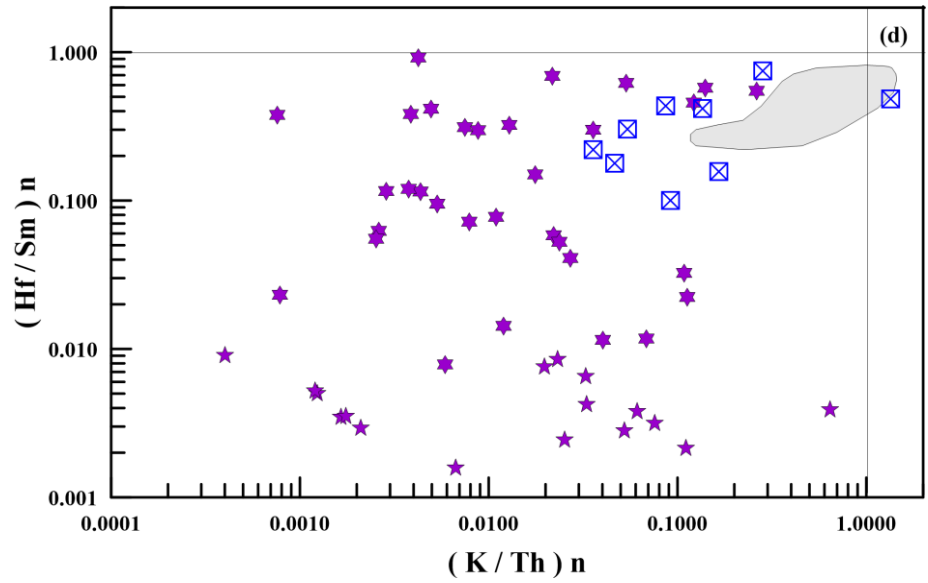

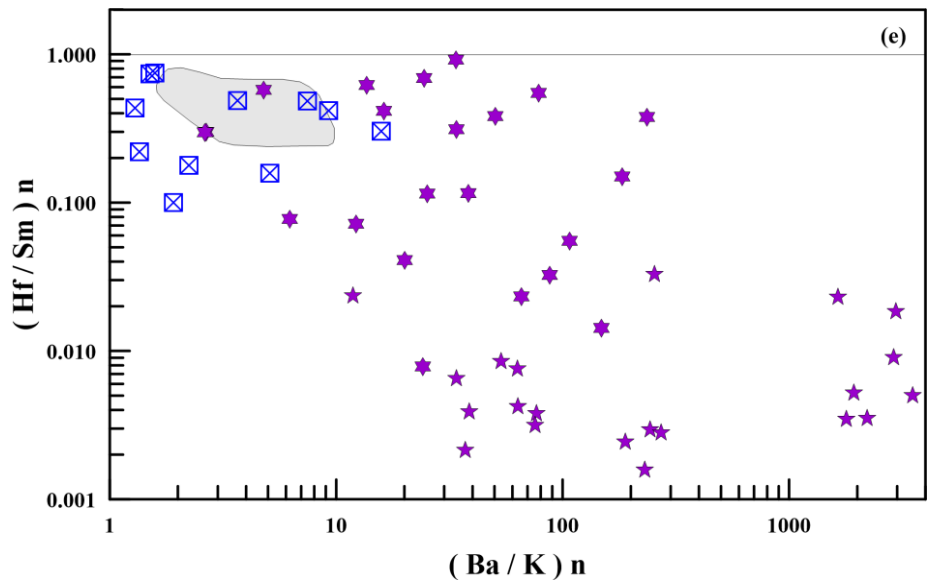

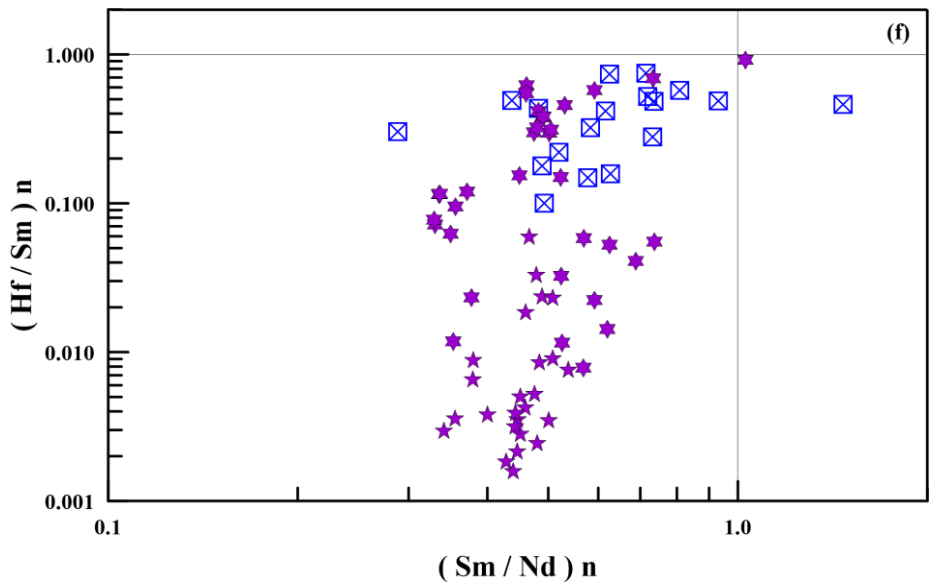

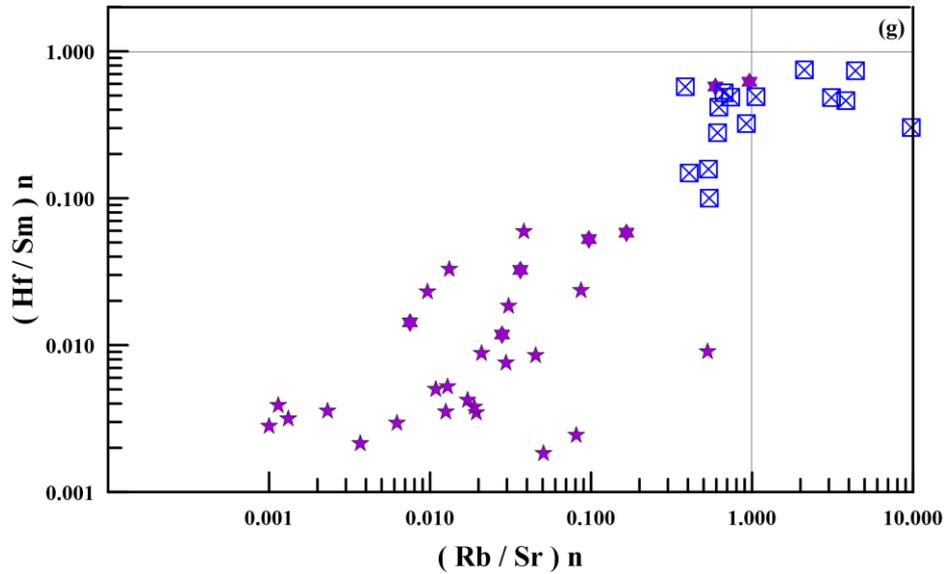

**Figure 4.** Composition of mantle xenoliths and carbonatite melt/magma composition (a) $(Hf/Sm)_n$ versus $(Ti/Eu)_n$; (b) $(Hf/Sm)_n$ versus $(La/K)_n$; (c) $(Hf/Sm)_n$ versus $Zr/Hf$; (d) $(Hf/Sm)_n$ versus $(K/Th)_n$; (e) $(Hf/Sm)_n$ versus $(Ba/K)_n$; (f) $(Hf/Sm)_n$ versus $(Sm/Nd)_n$; (g) $(Hf/Sm)_n$ versus $(Rb/Sr)_n$. 1 - composition of mantle xenoliths affected by carbonatite metasomatism; 2 - composition of continental carbonatites; 3 - oceanic carbonatite melts; 4 – pyroxenite xenoliths from Salt Lake Crater, Hawaii Islands; 5 – the rejuvenated-stage Hawaiian lavas and glasses. The database on composition of mantle peridotites affected by carbonatite metasomatism is after (Rudnick et al., 1993; Ionon et al., 1993; Yaxley et al., 1991). Data on extrusive carbonatite magma and melt composition are taken from Nelson et al. (1988), Gerlach et al. (1988), Woolley et al. (1991), Beccaluva et al. (1992), Tayoda et al. (1994) and Hoernle et al., (2002). Data on composition of pyroxenite xenoliths are from Frey et al. (1980). The data source is given in Supplementary Datasets 1, 2 and Table 1. Lines correspond to the composition of the primitive mantle according to Sun and McDonough (1989).

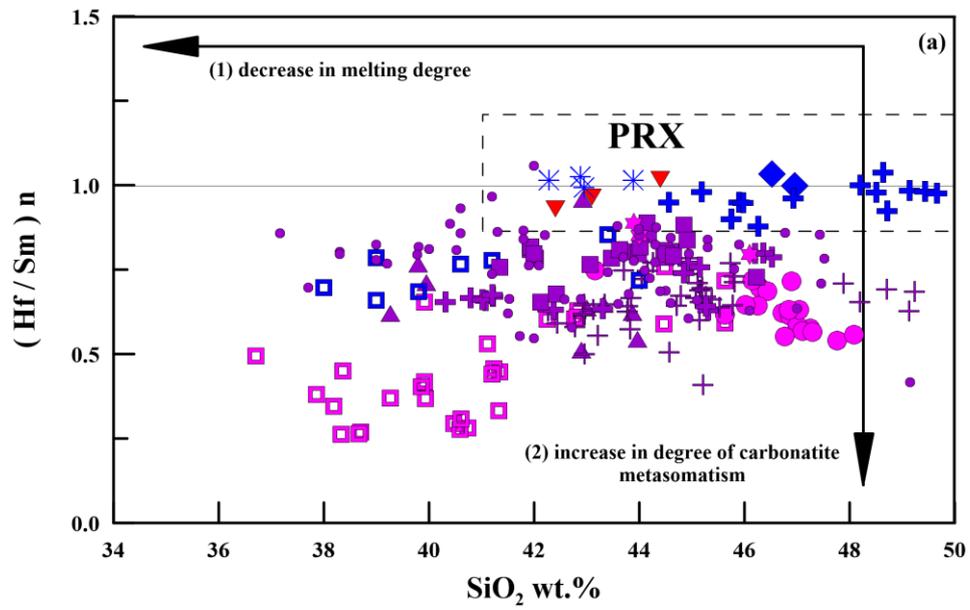
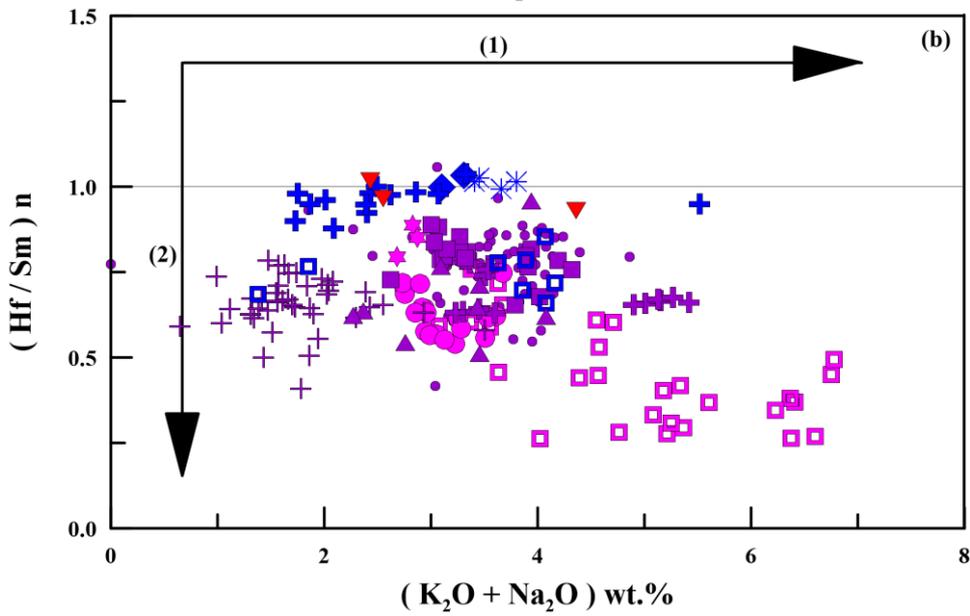
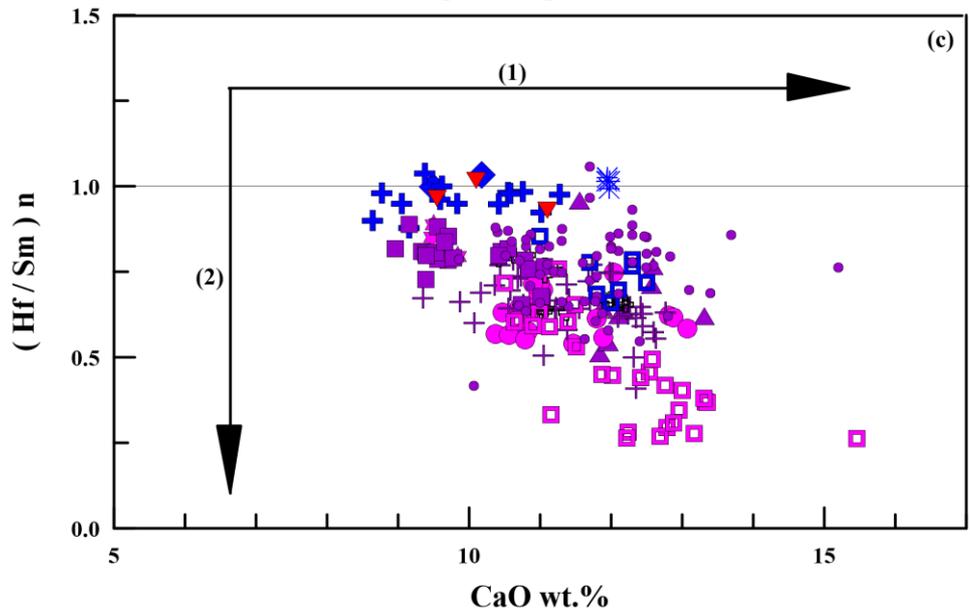

**Figure 5.** (a) (Hf/Sm)$_n$ versus SiO$_2$ contents versus; (b) (Hf/Sm)$_n$ versus (K$_2$O+Na$_2$O) contents (c) (Hf/Sm)$_n$ versus CaO contents in the Hawaiian lava series. The used data sets are given in Supplementary Dataset 1. (1) Trend of decreasing melting degree (for CaO contents the trend may also reflect the increasing degree of carbonatite metasomatism influence on mantle source); (2) trend of increasing degree of carbonatite metasomatism influence on mantle source is shown. Field shows suggested composition of melts derived from pyroxenite veins (PRX) of the Hawaiian mantle. Lines of constant (Hf/Sm)$_n$ correspond to the primitive mantle composition. The data sources are the same as in Figure 1.

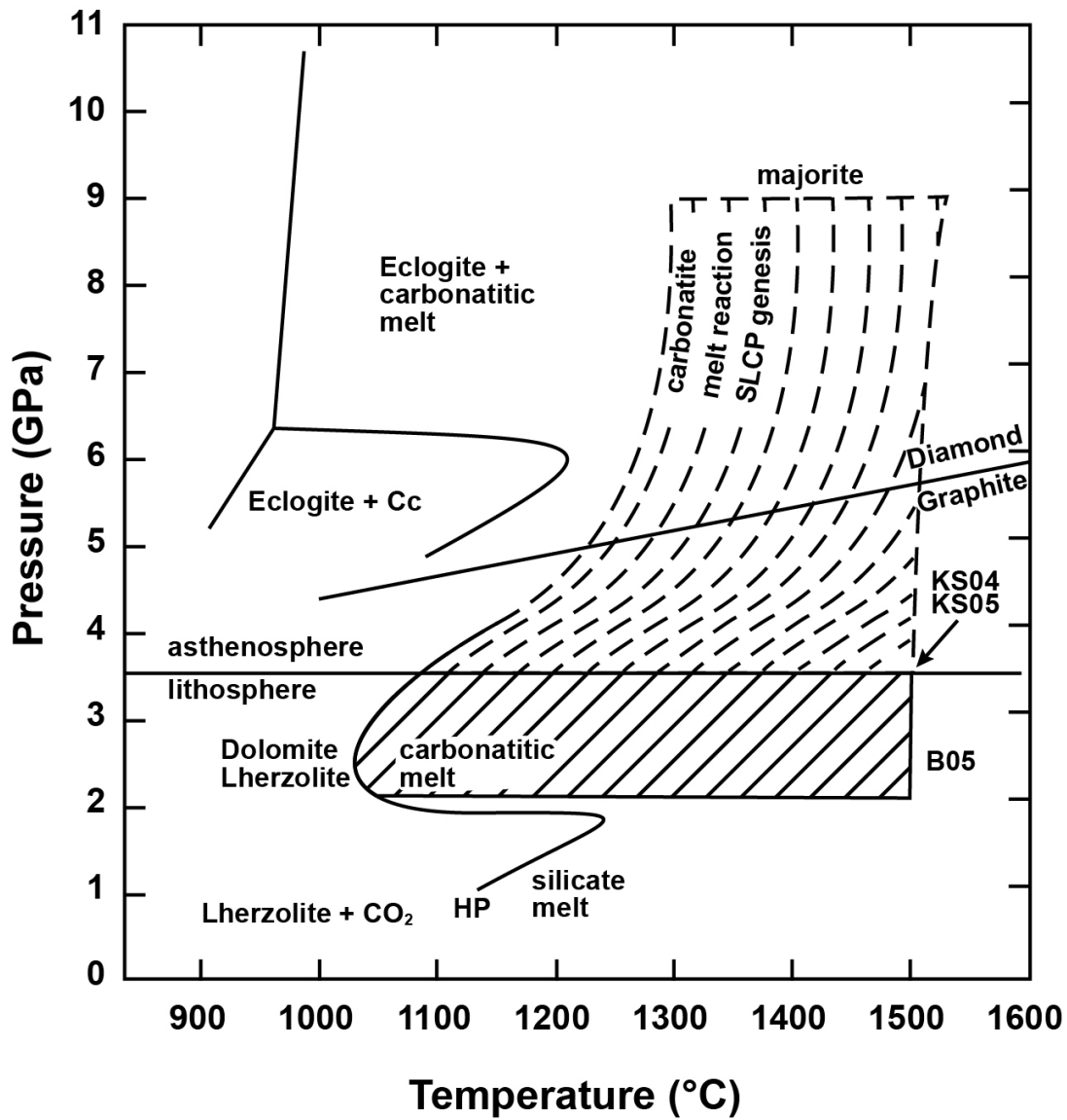

**Figure 6.** Model of the carbonated mantle melting beneath the Hawaiian plume. Our model suggests that the rejuvenated-stage Hawaiian lavas are generated due to partial melting of mantle lithospheric to asthenospheric peridotite source affected by carbonatite metasomatism at temperatures higher than 1100°C and pressures higher than 2 GPa. The source of the primary carbonatite melts may be observed in the Salt Lake Crater garnet pyroxenite (SLCP) xenoliths hosted by the Koolau volcano lavas on Oahu, Hawaii. The xenoliths demonstrate evidences for deep (up to 9 GPa) eclogite source generating the carbonatite melts within the Hawaiian plume. These carbonatite melts bear the Salt Lake Crater pyroxenite xenolith series to the surface and are responsible for the melt-rock reactions recorded in the xenoliths. HP means Hawaiian pyrolite, Cc - calcite. B05 means Bizimis et al (2005) and KS04 and KS05 is Keshlav & Sen (2004; 2005). The P-T diagram is adapted from Hammouda (2003) and modified.

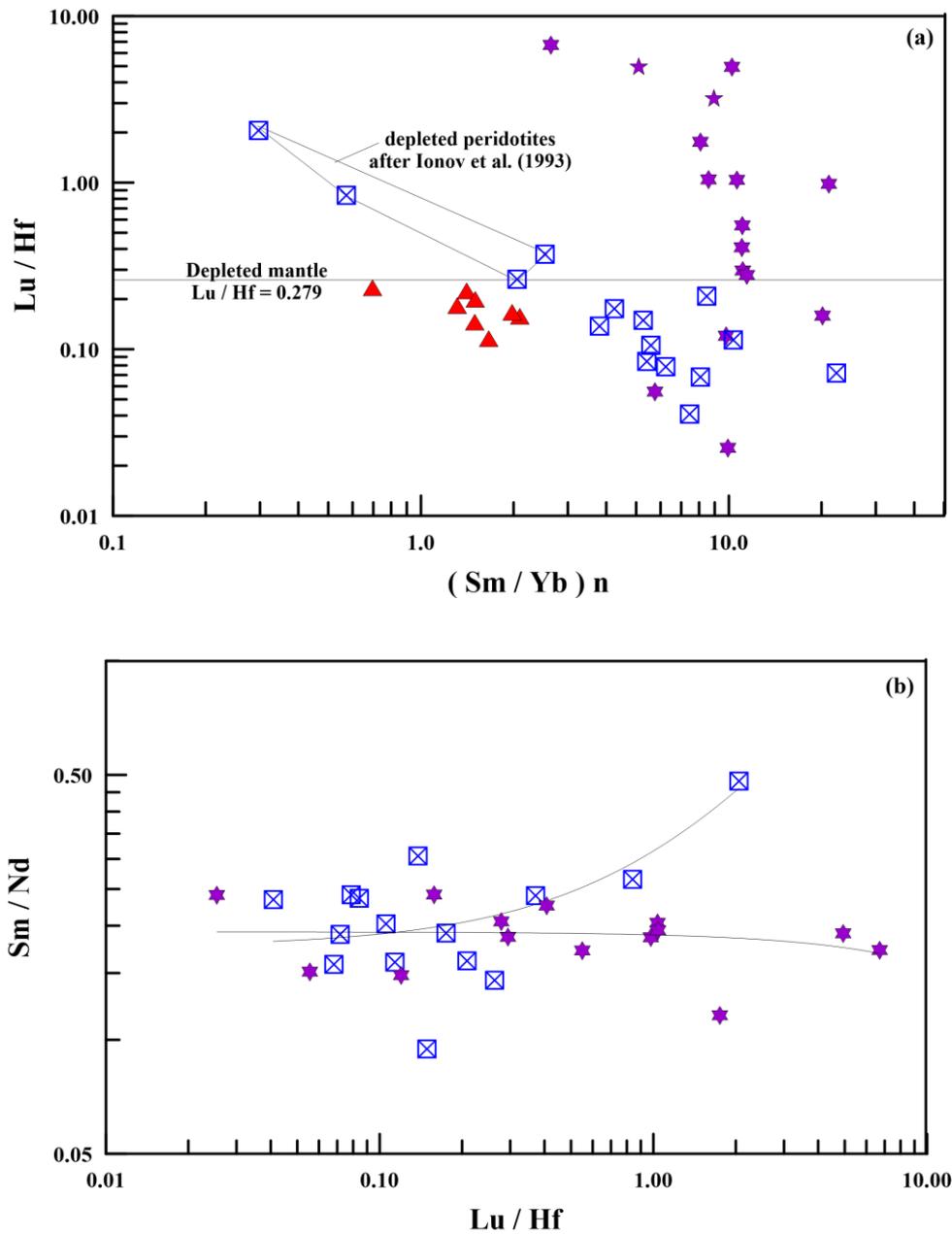

**Figure 7.** (a) (Hf/Sm)$_n$ versus (Hf/Yb)$_n$; (b) Lu/Hf versus (Hf/Yb)$_n$; (c) Lu/Hf versus (Sm/Yb)$_n$; (d) Sm/Nd versus Lu/Hf. Lu/Hf ratio for the depleted mantle is after Faure (1986). The used data sets are given in Supplementary Dataset 2. The data sources are the same as in Figure 4.

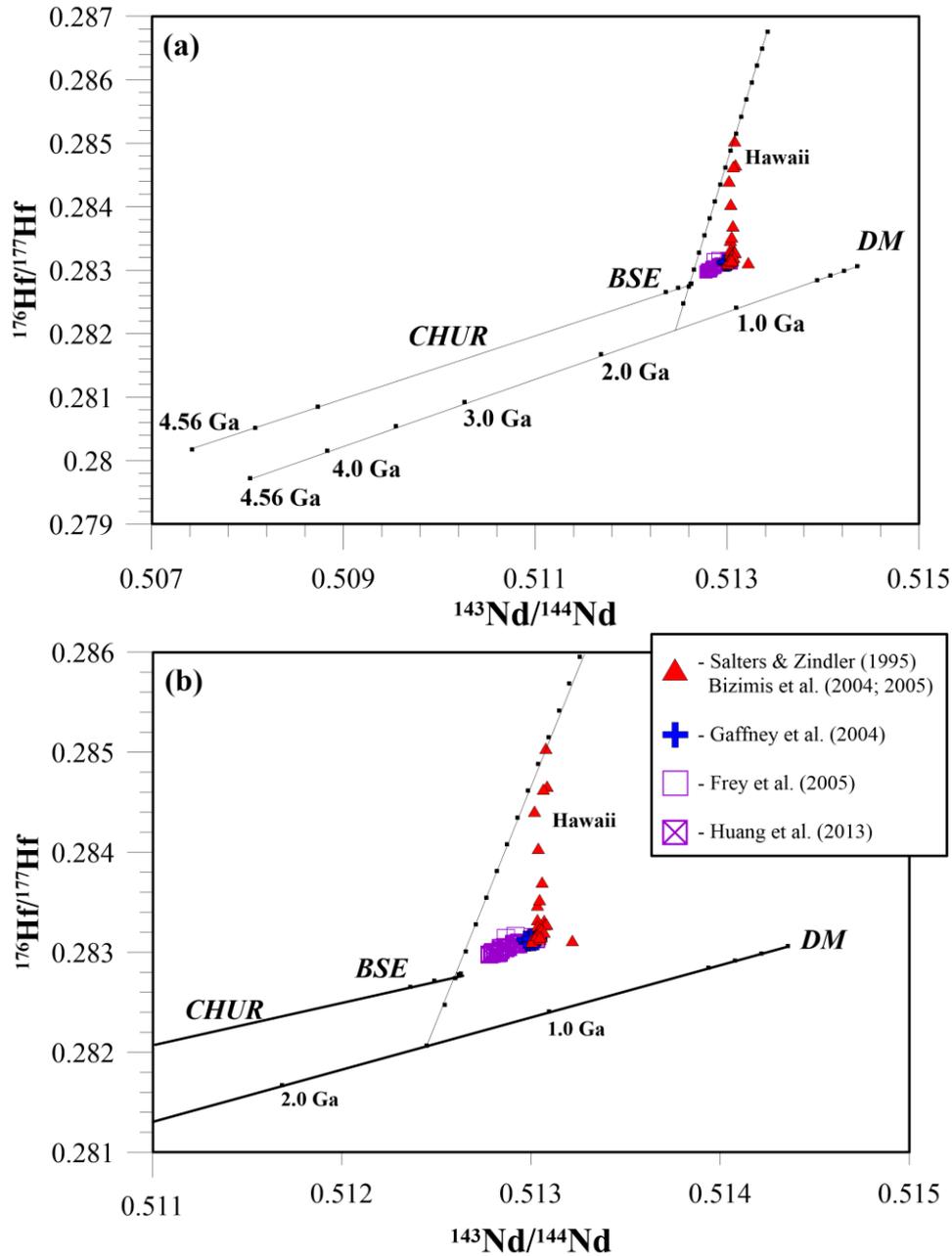

**Figure 8.** (a,b) $^{176}$Hf/$^{177}$Hf versus $^{143}$Nd/$^{144}$Nd. Model evolution of Hf-Nd isotope systematics of depleted mantle (DM) affected by carbonated metasomatism of 1 Ga event. The Nd-Hf isotopic ratios of the rejuvenated-stage lavas are after Frey et al. (2005), Gaffney et al. (2004) and Huang et al. (2013) and the Salt Lake Crater mantle xenoliths are after Salters & Zindler (1995) and Bizimis et al. (2004; 2005). Numbers correspond to model age of the depleted mantle (DM) reservoir. BSE is the Bulk Silicate Earth reservoir (Zindler & Hart, 1986). The model is calculated according to the initial chondritic uniform reservoir (CHUR) composition: ($^{176}$Hf/$^{177}$Hf)$_i$ = 0.279793 at 4.56 Ga and $^{176}$Lu/$^{177}$Hf = 0.0336 (Blichert-Toft & Albarède, 1997; Bouvier et al., 2008). The model is calculated according to DM composition: ($^{176}$Hf/$^{177}$Hf)$_i$ = 0.279718 at 4.56 Ga and $^{176}$Lu/$^{177}$Hf = 0.0384 (Griffin et al., 2000). The Nd isotope model is adapted from Faure (1986). The two-stage model suggests that carbonatite metasomatism might affect the DM reservoir during >1 Ga events. The data source is given in the Supplementary Model.

**Table 1.** Location, age and characteristics of the selected Hawaiian lava and glass samples

| Stage of magmatism | Location | Type of samples | Age | Reference |
|---|---|---|---|---|
| Post-erosional or rejuvenated* | Honolulu volcano, Oahu island | nepheline melilitite; nephelinite; nephelinite with rare melilite; basanite; alkali olivine basalt | < 0.58 Ma | Clague & Frey, 1982 |
| Preshield** | Loihi | basanite and alkali basalt | 5 ± 4 to 102 ± 13 ka | Frey & Clague, 1983 |
| Postshield and rejuvenated | Kauai island | alkalic basalt, basanite, nephelinite, and nepheline melilitite | between 3.7 and 0.52 Ma | Clague & Dalrymple 1988 |
| Postshield*** | Haleakala volcano, Maui island | alkalic basalts | 0.5 – 0.1 Ma | Chen et al. 1991 |
| Post-erosional or rejuvenated | Koloa volcano, Kauai island | melilitite; nephelinite; basanite; alkali olivine basalt | a peak in the activity at 1.2 Ma | Maaloe et al. 1992 |
| Rejuvenated | Kauai island | basanites, alkali basalts | from about 3.5 to 0.5 Ma | Reiners & Nelson 1998 |
| Rejuvenated | North Arch volcanic field | alkalic basalt to nephelinite | 1.15 – 0.5 Ma | Frey et al., 2000 |
| Rejuvenated | submarine Wailau landslide, Molokai | basanites and alkaline basalts | between 1.5 and 1.4 Ma | Clague & Moore, 2002 |
| Postshield and rejuvenated | West Maui | olivine basalts | 0.6 – 0.4 Ma | Gaffney et al. 2004 |
| Rejuvenated* | Honolulu volcano, Oahu island | alkalic basalt to nephelinite | < 0.58 Ma | Clague et al. 2006 |
| Rejuvenated | Kiekie basalt, Niihau island | basalt glasses | 3.5 – 0.35 Ma | Dixon et al., 2008 |
| Rejuvenated | Kauai island | transitional basalt to nephelinite | 2.3 – 0.3 Ma | Cousens & Clague, 2015 |
| Postshield*** | Haleakala, Maui island | basanites | < 0.15 Ma | Phillips et al., 2016 |

\* Lanphere and Dalrymple (1980);
\*\* Guillou et al. (1997);
\*\*\* Sherrod et al. (2003).

**Table 2.** Trace element partition coefficients used in the modeling

| | Ol | | Opx | | Cpx | | Spl | | Grnt | | Amph | | Phl | |
|---|---|---|---|---|---|---|---|---|---|---|---|---|---|---|
| Zr | 0.0005 KE [a] 0.0025 R | | 0.014 | KE | 0.089 | S | 0.001 | R | 0.27 | J | 0.127 | L | 0.017 | L |
| | 0.0003 FJ | | 0.027 | R | 0.1234 | HD | 0.07 | KE | 0.3 | KE | 0.23 | BR | 0.13 | A |
| | | | 0.032 | GR | 0.1280 | J | | | 2.12 | H | 0.18-0.33 | A | 0.23 | BR |
| | | | 0.0099 | AGR | 0.164 | H | | | 0.47 | AGR | | | 0.008 | GR |
| | | | | | 0.18 | R | | | | | | | | |
| | | | | | 0.27 | BL | | | | | | | | |
| Hf | 0.0028 | R | 0.025 | R | 0.10-0.20 | T | 0.001 | R | 0.24 | J | 0.33 | L | 0.19 | L |
| | 0.001-0.004 | D | 0.04 | D | 0.179 | S | 0.003 | E | 0.62 | T | 0.45 | BR | 0.45 | BR |
| | 0.0008 | FJ | 0.017 | AGR | 0.19 | F | | | 1.22 | H | | | 0.091 | GR |
| | | | 0.06 | GR | 0.23 | J | | | | | | | | |
| | | | | | 0.256 | HD | | | | | | | | |
| | | | | | 0.29 | H | | | | | | | | |
| | | | | | 0.3 | R | | | | | | | | |
| | | | | | 0.34 - 0.38 | W | | | | | | | | |
| | | | | | 0.36 - 0.46 | D | | | | | | | | |
| | | | | | 0.55 | BL | | | | | | | | |
| | | | | | 0.13 | E | | | | | | | | |

**Table 2** (continued):

| | Ol | Opx | Cpx | Spl | Grnt | Amph | Phl |
|---|---|---|---|---|---|---|---|
| Sm | 0.0007 KE<br>0.0025 R | 0.014 R<br>0.02 KE<br>0.011 AGR<br>0.015 GR | 0.086-0.22 T<br>0.201 S<br>0.281 J<br>0.291 HD<br>0.33 H<br>0.35 F<br>0.4 R<br>0.67 BL<br>0.31 E<br>0.21 GR | 0.0006 KE<br>0.0006 R | 0.23 T<br>0.25 J<br>0.5 KE<br>0.101 GR | 0.66 BR | 0.27 G<br>0.66 BR<br>0.017 GR |
| La | 7E-06 KE<br>4.5E-04 R | 0.0005 KE<br>0.0025 R | 0.03 S<br>0.04-0.28 AG<br>0.0536 HD<br>0.06 R<br>0.089 BL | 0.0001 R<br>0.0006 KE | 0.001 KE<br>0.0016 J<br>0.007 L | 0.055 L | 0.06 G |
| K | 1E-09 KE | 1E-05 KE | 0.0072 HD | 0 KE | 1E-05 KE | 0.58 L | 3.67 L |
| Ba | 1E-09 KE * | 1E-05 KE | 0.00068 HD | 0 KE | 1E-05 KE | 0.16 L | 3.68 L |
| Th | 7E-06-1E-05 B<br>1.3E-04 R | 2E-05-3E-05 B<br>12.5E-05 R | 1.5E-05 R<br>1.3E-03-2.1E-03 B<br>0.0070 T<br>0.0086 F<br>0.014 H | 1E-05 R<br>0.0014 H<br>0.0036 T | 2.1$^E$-03 B | 0.0039 L<br>0.017 BR | 0.0014 L |
| Nb | 0.0001 KE<br>0.0017 R | 0.0025 R<br>0.003 KE | 0.003 S<br>0.0077 HD<br>0.008 R<br>0.020 BL | 0.01 KE<br>0.07 R | 0.0042 J<br>0.01 KE | 0.159 L<br>0.20 BR | 0.088 L |
| Ti | 0.015 KE | 0.14 KE | 0.273 S<br>0.347 J<br>0.35-0.43 D<br>0.384 HD | 0.15 KE | 0.28 J<br>0.6 KE | 0.95 A<br>1.29 L | 0.98 A<br>1.77 L |
| Eu | 9.5E-04 KE<br>0.0029 R | 0.0185 R<br>0.03 KE | 0.35 HD<br>0.38 BL<br>0.46 R | 6E-04 KE<br>9E-04 R | 1 KE | 0.88 IF | 0.029 IF |

$^a$ Trace element partition coefficients ($K_d = C_i^{mineral} / C_i^{melt}$) between the main mantle minerals and basaltic or basanitic melt are according to A - Adam et al. (1993); AG - Adam and Green (1994); AGR – Adam and Green (2006); B – Beattie (1993); BL - Blundy et al. (1998); BR - Brenan et al. (1995); D – Dunn (1987); E – Elkins et al. (2008); F - Falloon et al. (1988); FJ - Foley and Jenner (2004) ; GA – Gaetani et al. (2003); G – Green (1994); GR - Green et al. (2000); H - Hauri et al. (1994); HD - Hart and Dunn (1993); Irving and Frey (1984); J – Johnson (1994); KE - Kelemen et al. (1993); L - LaTourette et al. (1995); R – Remaidi (1993); S - Skulski et al. (1994); T – Takahashi (1986); W - Watson et al. (1987). Underlined values were used for calculations. Ol, Opx, Cpx, Spl, Grnt, Amph Phl denotes olivine, orthopyroxene, clinopyroxene, spinel, garnet, amphibole, phlogopite.

* $K_d^{Ba}$ for Ol, Opx, Spl and Grnt olivine, orthopyroxene, clinopyroxene, spinel, garnet, amphibole, phlogopite are suggested as equal to those of $K_d^K$.

# APPENDIX

**Estimation of trace element ratios in the partial melts relatively to those of amphibole-bearing (A) and phlogopite-bearing (P), spinel (S) and garnet (G) lherzolites sources**

The values of X/Y fractionation in melt relatively to X/Y in mantle source (**Fig. 1**) were calculated using the equation of batch melting from Shaw (1970): $([X]_{melt}/[Y]_{melt})/([X]_{source}/[Y]_{source}) = [D_o^Y + f(1-D_o^Y)] / [D_o^X + f(1-D_o^X)]$, where f is degree of melting of mantle source, $D_o$ corresponds to the bulk distribution coefficients, [X] and [Y] are concentrations of X and Y in the elemental ratio of X and Y elements.

$D_o$ for A and P lherzolites have been calculated using $K_d$ for Ol, Opx, Cpx, Amph and Phl (Table 1) and mineral composition of A and P lherzolites taken according to McKenzie and O'Nions (1991): 59.9% Ol, 24.4% Opx, 3.8% Cpx, 11.6 % Amph (Phl): $D_o$ for Zr (0.0232 (A) and 0.0104 (P)), Hf (0.0559 and 0.0396), Sm (0.093 and 0.0477), La (0.0085 and 0.0091), K (0.0676 and 0.426), Ba (0.0186 and 0.4269), Nb (0.0195 and 0.0113), Th (0.0006 and 0.0003), Ti (0.2064 and 0.2619) and Eu (0.1234 and 0.0246).

$D_o$ for S and G lherzolites have been calculated using $K_d$ for Ol, Opx, Cpx, Spl and Grnt (Table 1), mineral composition of lherzolites taken according to McKenzie and O'Nions (1991): S (57.8% Ol, 27% Opx, 11.9% Cpx, 3.3% Spl) and G (59.8% Ol, 21.1% Opx, 7.6% Cpx and 11.5% Grnt): $D_o$ for Zr (0.021 (S) and 0.047 (G)), Hf (0.039 and 0.054), Sm (0.0405 and 0.0843), La (0.0065 and 0.0043), K (0.00086 and 0.0006), Ba (8.4E-05 and 5E-05), Nb (0.0021 and 0.0024), Th (0.00011 and 0.0003), Ti (0.0928 and 0.134), Eu (0.0503 and 0.1485).

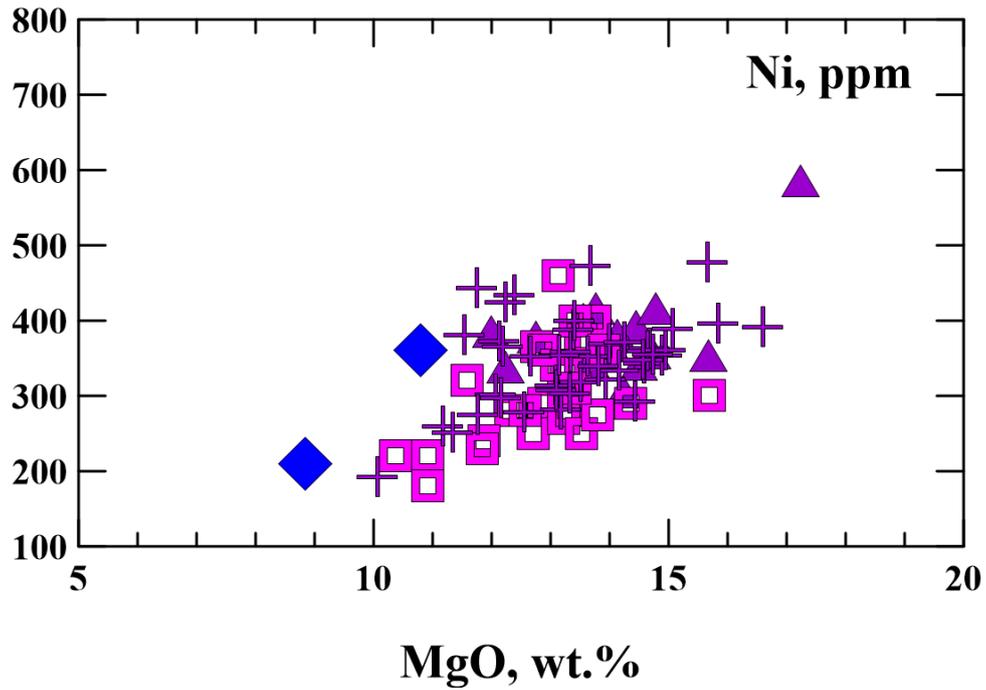

**Figure A1.** Ni (in ppm) versus MgO (wt.%) contents in the Hawaiian alkaline lavas examined in this work. The data source is the same as in Figure 1.

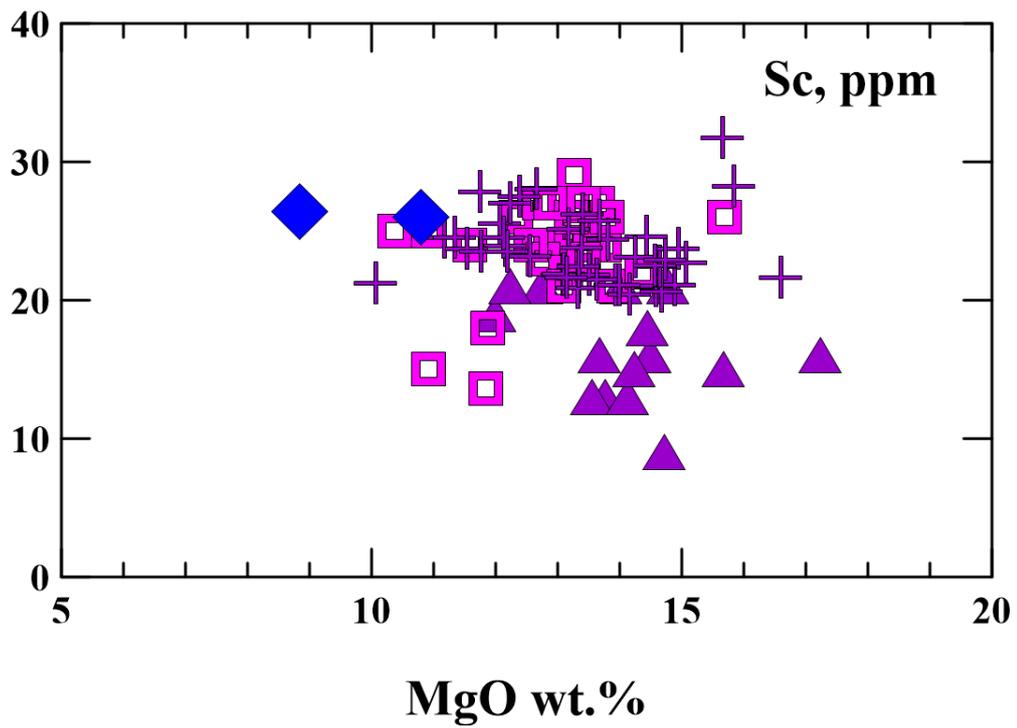

**Figure A2.** Sc (in ppm) versus MgO (wt.%) contents in the Hawaiian alkaline lavas examined in this work. The data source is the same as in Figure 1.